\documentclass[letterpaper]{article}
\usepackage[preprint]{aaai2027}
\usepackage[hyphens]{url}
\usepackage{graphicx}
\usepackage{natbib}
\usepackage{caption}
\usepackage{amssymb}
\usepackage{amsmath}
\usepackage{amsthm}
\usepackage{algorithm}
\usepackage{algpseudocode}
\usepackage{booktabs}

\newtheorem{theorem}{Theorem}
\newtheorem{lemma}{Lemma}
\newtheorem{corollary}{Corollary}
\newtheorem{proposition}{Proposition}
\theoremstyle{remark}

\title{Fast Onboard Mars Ascent Parameter Determination via Parallel Refinement of Learned Initial Guesses}

\author{
    Yushu Chen\textsuperscript{\rm 1},
    Hongkun Yu\textsuperscript{\rm 1},
    Wenlai Zhao\textsuperscript{\rm 1},
    Kangyi Zhan\textsuperscript{\rm 2},
    Haipeng Chen\textsuperscript{\rm 2},
    Tao Zhang\textsuperscript{\rm 3},
    Wusheng Zhang\textsuperscript{\rm 1},
    Guangwen Yang\textsuperscript{\rm 1}\corresponding
}
\affiliations{
    \textsuperscript{\rm 1}Department of Computer Science and Technology, Tsinghua University, Beijing 100086, China;\\
    also at National Supercomputing Center in Wuxi\\
    \textsuperscript{\rm 2}Shanghai Aerospace Control Technology Institute, Shanghai 201109, China\\
    \textsuperscript{\rm 3}School of Computer Science, Beijing University of Posts and Telecommunications, Beijing 100876, China
}

\begin{document}

\maketitle

\begin{abstract}
Onboard ascent parameter determination is essential for a Mars ascent vehicle because post-landing launch conditions are unknown a priori and Earth--Mars communication is delayed. While radiation-hardened flight processors have slow individual cores, modern aerospace architectures increasingly offer multi-core parallelism. However, sequential quadratic programming (SQP) requires hundreds of trajectory simulations across dozens of serial steps, whereas standalone neural networks fail to meet strict orbit-insertion tolerances. We propose a fast onboard solver based on a predict-then-refine paradigm, pairing an offline-trained residual network for near-optimal parameter prediction with an online boundary-frozen damped singular value decomposition (SVD) refinement that restores the circular-orbit terminal conditions via event-aligned simulations. Each refinement iteration executes in just two parallel rounds of mutually independent simulations. We prove local convergence under bounded prediction error and show that propellant suboptimality scales quadratically with the prediction error when inequality constraints are inactive (and at most linearly otherwise). Across 3,000 test missions, the method achieved 100\% compliance with all terminal-orbit and dynamic-pressure constraints, keeping second-stage propellant consumption within 0.0065\% of the KKT-optimal SQP reference in 95\% of cases. Under identical operational stopping tolerances, relative to cold-started early-exit SLSQP, average simulations dropped from 271 to 32 and sequential rounds from 32.8 to 3.3, yielding an eight-core server speedup of $11.8\times$ ($2.2\times$ over ResNet-warm-started early-exit SLSQP and $2.3\times$ over Newton--GMRES) and an $8.9\times$ serial speedup on a bare-metal aerospace processor.
\end{abstract}

\section{Introduction}\label{sec:intro}

Delivering a payload from the Martian surface into a target circular orbit requires a Mars ascent vehicle (MAV) to determine key ascent parameters---including launch azimuth, pitch profile parameters, and burn and coast durations---that depend on the actual landing site, vehicle state, and target orbit known accurately only after touchdown. Because Earth--Mars communication involves ten-minute-scale one-way delays and intermittent contact windows, ground-based computation and uplink cannot timely accommodate post-landing states or mission updates, making autonomous onboard parameter determination essential prior to lift-off. Moreover, tight rendezvous timing constraints, updated environmental or target parameters, and potential retargeting contingencies leave only a brief pre-ignition window for onboard generation and rapid re-computation. This pre-launch task is complementary to closed-loop guidance laws such as the iterative guidance mode (IGM) \cite{chandler1967igm} and powered explicit guidance (PEG) \cite{jaggers1977peg}, which compute in-flight steering corrections around the nominal ascent established before launch.

Onboard computation, however, is severely constrained by flight hardware. Radiation-hardened processors operate at low clock frequencies and deliver far lower instruction throughput per cycle than commercial counterparts \cite{dueri2017customized,lentaris2018hpec}. With multi-core radiation-hardened architectures now emerging \cite{rogers2016maestro,tambara2019gr740,yun2025multicore}, additional onboard compute comes primarily from core-level parallelism rather than faster single-core execution. Conventional ascent trajectory generation relies on numerical optimization---such as direct collocation, sequential convex programming (SCP), and sequential quadratic programming (SQP) \cite{liu2017survey,malyuta2021advances,benedikter2021ascent,ma2022mcpi,reducedspace2025}---whose iterations are inherently serial. In our eight-parameter ascent problem, SciPy's SLSQP solver averages 396 trajectory simulations across 67 sequential rounds per mission; because concurrency is confined within each finite-difference batch, additional cores cannot compress this serial depth.

Machine learning shifts the computational burden offline, allowing neural networks to predict control parameters in a single forward pass \cite{sanchez2018landing,rubinsztejn2020missedthrust,wang2022ascent,fixedwing2026}. Yet approximation errors prevent direct orbit insertion: in our benchmark, a residual network (ResNet) \cite{he2016resnet} trained offline on 24,000 SQP solutions achieves a median semi-major axis error of 1.2~km, with zero of 3,000 test predictions meeting the strict terminal tolerances (30~m in semi-major axis, $10^{-6}$ in eccentricity, $10^{-6}$~rad in inclination).

A natural remedy is to warm-start a numerical optimizer with the learned prediction, a strategy explored in SCP \cite{banerjee2020warmstart,dambrosio2026sequence,takubo2026transformer,banerjee2025iss}, indirect shooting \cite{shi2022indirect,initialcostates2024}, constraint-informed primal--dual learning \cite{briden2025constraint}, nonparametric trajectory libraries \cite{lembono2020memory}, and fault-tolerant MAV replanning \cite{li2025marswarmstart}. However, these frameworks still execute online constrained optimization---initializing quasi-Newton Hessian approximations from scratch and solving coupled quadratic subproblems---so a multi-step outer iteration chain persists. In our experiments, even when warm-started from the ResNet prediction and terminated immediately upon entering the target orbital tolerance box, SLSQP still averages 2.66 outer iterations and 46.2 simulations across 6.3 sequential rounds. This motivates a key question: given a near-optimal yet imprecise learned prediction, can an onboard solver rapidly restore terminal feasibility on multi-core hardware at negligible propellant cost?

This paper answers this question by departing from online optimization: an offline-trained ResNet predicts the ascent parameters near the propellant-optimal solution, and an online boundary-frozen damped singular value decomposition (SVD) refinement restores terminal feasibility on the four orbital equations. Bypassing objective gradients and quadratic subproblems, each iteration executes in just two parallel simulation rounds (central-difference perturbations and four concurrent candidate step lengths), while the minimum-norm SVD update suppresses drift along the feasible manifold to preserve propellant near-optimality.

The main contributions of this work are summarized as follows:
\begin{enumerate}
\item \textbf{Optimization-free onboard paradigm:} Replacing serial online optimization with offline ResNet prediction and online feasibility restoration cuts average simulations from 271 to 32 and sequential rounds from 32.8 to 3.3 under identical operational tolerances (and from 396 simulations and 67 rounds vs.\ full-KKT SLSQP), achieving 100\% compliance and propellant within 0.0065\% of the SQP reference in 95\% of 3,000 test missions.
\item \textbf{Parallel boundary-frozen SVD refinement:} Freezing near-bound parameters prior to central differencing avoids clipping distortion and dispatches Jacobian perturbations ($\le 16$) and trial steps ($4$) in two parallel rounds per iteration.
\item \textbf{Theoretical convergence and suboptimality guarantees:} We establish an explicit initial-error threshold for local convergence and prove an exact KKT decomposition showing that propellant penalty scales quadratically with network error when inequality constraints are inactive or locked, and at most linearly otherwise.
\item \textbf{Multi-core server and flight-processor benchmarking:} Evaluated in C++ on an eight-core server and a bare-metal aerospace processor against equal-tolerance early-exit SLSQP and Newton--GMRES baselines, the method achieves $11.8\times$ eight-core speedup over cold early-exit SLSQP ($2.2\times$ over warm early-exit SLSQP; $2.3\times$ over Newton--GMRES) and $8.9\times$ serial speedup on the flight processor ($1.5\times$ and $1.3\times$, respectively).
\end{enumerate}

The remainder of this paper is organized as follows. Section~\ref{sec:problem} formulates the ascent parameter determination problem and defines computational cost in terms of total simulations and sequential rounds. Section~\ref{sec:method} details the ResNet predictor and the parallel boundary-frozen damped SVD refinement. Section~\ref{sec:theory} establishes theoretical guarantees for convergence and propellant suboptimality. Section~\ref{sec:experiments} presents numerical benchmarks and aerospace processor timing results, and Section~\ref{sec:conclusion} concludes the study.

\section{Problem Formulation}\label{sec:problem}

\subsection{Ascent trajectory model and event handling}\label{sec:model}

We consider a two-stage Mars ascent vehicle modelled as a point mass. Its position $\mathbf r$ and velocity $\mathbf v$ are expressed in a Mars-centred inertial frame, and its mass is $m$. The equations of motion are
\begin{equation}
\dot{\mathbf r}=\mathbf v,\quad
\dot{\mathbf v}=\mathbf g(\mathbf r)+\frac{1}{m}\left(\mathbf F_{T}+\mathbf F_{A}\right),\quad
\dot m=-\dot m_k ,
\label{eq:eom}
\end{equation}
where $\mathbf g$ is the gravitational acceleration of an oblate Mars including the $J_2$ term, $\mathbf F_T$ is the thrust of the active stage $k$, $\mathbf F_A$ is the aerodynamic force (axial drag and a normal force proportional to the angle of attack), and $\dot m_k$ is the mass flow rate of stage $k$. The atmospheric density follows an exponential pressure model and an altitude-dependent temperature profile, and the aerodynamic forces use the velocity relative to the rotating atmosphere. The first-stage thrust is $\dot m_1u_1+S_e(p_0-p)$, where $u_1$ is the effective exhaust velocity, $S_e$ the nozzle exit area, $p$ the ambient pressure and $p_0$ the surface pressure; the second-stage thrust is constant. The physical constants and vehicle data are listed in Table~\ref{tab:model}; the vehicle data are representative values for a small two-stage ascent vehicle and do not describe a specific vehicle.

\begin{table}[t]
\centering
\caption{Mars model and vehicle data.}\label{tab:model}
\footnotesize
\setlength{\tabcolsep}{2.5pt}
\begin{tabular}{lll}
\hline
Quantity & Value & Unit\\
\hline
\multicolumn{3}{l}{\emph{Mars}}\\
Equatorial radius $R_{\mathrm{ref}}$ & 3397.2 & km\\
Polar radius & 3375.5 & km\\
Mean radius & 3396.19 & km\\
Gravitational parameter $\mu$ & $4.2828\times10^{13}$ & m$^3$/s$^2$\\
Second zonal harmonic $J_2$ & $1.96\times10^{-3}$ & --\\
Rotation rate $\omega$ & $7.090\times10^{-5}$ & rad/s\\
Surface pressure $p_0$ & 559.35 & Pa\\
Pressure at altitude $h$ (m) & $p_0e^{-1.05\times10^{-4}h}$ & Pa\\
\multicolumn{3}{l}{\emph{Vehicle}}\\
Lift-off mass & 400 & kg\\
Second-stage mass & 150 & kg\\
Propellant mass $m_{p1}$ / $m_{p2}$ & 220 / 80 & kg\\
Nominal 1st-stage flow rate $\dot m_{1,\mathrm{nom}}$ & 2.2 & kg/s\\
Second-stage flow rate $\dot m_2$ & 0.34375 & kg/s\\
First-stage exhaust velocity $u_1$ & 3000 & m/s\\
First-stage nozzle exit area $S_e$ & 0.18 & m$^2$\\
Second-stage exhaust velocity & 3200 & m/s\\
Second-stage thrust & 1100 & N\\
Reference area & 0.2 & m$^2$\\
Axial force coefficient & 0.50 & --\\
Normal force coefficient slope & 0.12 & 1/deg\\
\multicolumn{3}{l}{\emph{Timeline}}\\
Attitude hold after lift-off & 3 & s\\
Attitude hold at sep.\ and before cutoff & 3 & s\\
Stage separation time & 220 & s\\
Launch longitude & 0 & deg\\
\hline
\end{tabular}
\end{table}

The flight comprises six phases: (1)~a 3~s attitude hold at initial pitch $\phi_1$ after lift-off; (2)~first-stage powered flight with a prescribed angle-of-attack profile of peak $\alpha_m$; (3)~first-stage burnout at $T_{b1}=m_{p1}/\dot m_1$, followed by coast and fixed-time stage separation; (4)~second-stage ignition at $T_{b1}+t_c$, after interpolating attitude over the coast duration $t_c$ to $(\phi_2,\psi_2)$; (5)~second-stage powered flight of duration $T_2$ with linear pitch rate $\dot\phi_2$; and (6)~cutoff and terminal-state evaluation. The first-stage flow rate is $\dot m_1=\kappa\,\dot m_{1,\mathrm{nom}}$ for throttle factor $\kappa$, and $\dot m_2$ is constant, so second-stage propellant consumption is $\dot m_2T_2$.

The trajectory switches its dynamics or control law at the phase boundaries above, most of which depend on the ascent parameters (e.g., $t_c$ and $T_2$). To prevent fixed time-grid stepping from introducing artificial piecewise discontinuities into finite-difference derivatives, we split integration at every switching instant and apply a fourth-order Runge--Kutta (RK4) scheme within each segment at a 0.025~s step, evaluating $q_{b1}$ at the exact burnout instant. Across 24 validation missions, this 0.025~s event-aligned scheme differs from a 0.00625~s reference by at most 0.36\% of the accuracy tolerances of Section~\ref{sec:params} (at most 1.3~cm in semi-major axis) and $<10^{-9}$~Pa in burnout dynamic pressure, with smooth perturbation responses across grid points in all 72 test cases. All reported solutions are re-verified with the 0.00625~s step.

We denote one simulation of this model by
\begin{equation}
\left(\mathbf r_f,\mathbf v_f,\ q_{b1}\right)=\mathcal S(\boldsymbol\theta,\mathbf x),
\end{equation}
which maps the mission conditions $\boldsymbol\theta$ and the ascent parameters $\mathbf x$ to the terminal state $(\mathbf r_f,\mathbf v_f)$ and the dynamic pressure $q_{b1}$ at first-stage burnout. Each evaluation of $\mathcal S$ integrates the complete ascent trajectory.

\subsection{Ascent parameters, terminal conditions and constraints}\label{sec:params}

\paragraph{Mission conditions} A mission is described by
\begin{equation}
\boldsymbol\theta=\left(B_0,\ H_0,\ \kappa,\ a^\star,\ e^\star,\ i^\star\right),
\end{equation}
where $B_0$ and $H_0$ are the latitude and altitude of the launch site, $\kappa$ is the first-stage throttle factor, and $a^\star$, $e^\star$ and $i^\star$ are the semi-major axis, eccentricity and inclination of the target orbit. This paper considers circular target orbits, $e^\star=0$, with $a^\star=3{,}696{,}190$~m (300~km above the mean radius of Mars). The mission domain is $B_0\in[23^\circ,27^\circ]$, $H_0\in[0,6000]$~m, $\kappa\in[0.7,1.3]$ and $i^\star\in[30^\circ,60^\circ]$. The launch longitude is set to $0^\circ$.

\paragraph{Ascent parameters} The ascent parameter vector is
\begin{equation}
\mathbf x=\left(A_0,\ \phi_1,\ \phi_2,\ \psi_2,\ \dot\phi_2,\ t_c,\ T_2,\ \alpha_m\right)\in\mathbb R^8 .
\end{equation}
Their meanings and ranges are given in Table~\ref{tab:params}. The launch azimuth $A_0$ and the second-stage yaw $\psi_2$ are periodic; differences in these two angles are taken as the shortest angular distance. For a given target inclination, problem~(P) below has two families of local solutions, with launch azimuths near $\pi/2-i^\star$ (north-east) and near the mirror value $\pi/2+i^\star$ (south-east). We consider the second family, $A_0\in(\pi/2,\pi)$. The other six parameters are bounded. We normalize each parameter by its half range,
\begin{equation}
\begin{aligned}
\mathbf z&=\left(\mathbf x-\mathbf c_x\right)\oslash\mathbf s_x,\\
\mathbf c_x&=\tfrac12\left(\mathbf x^{up}+\mathbf x^{lo}\right),\quad
\mathbf s_x=\tfrac12\left(\mathbf x^{up}-\mathbf x^{lo}\right),
\end{aligned}
\label{eq:normalize}
\end{equation}
where $\oslash$ denotes element-wise division, and $\mathbf x^{lo}$ and $\mathbf x^{up}$ denote the lower and upper parameter bounds (or period endpoints for periodic angles). Each bounded coordinate then satisfies $-1\le z_j\le1$. The index set of the bounded coordinates is denoted by $\mathcal N$.

\begin{table}[t]
\centering
\caption{Ascent parameters and their ranges.}\label{tab:params}
\footnotesize
\setlength{\tabcolsep}{2pt}
\begin{tabular}{llll}
\hline
Symbol & Meaning & Range & Unit\\
\hline
$A_0$ & launch azimuth & $[0,2\pi)$, periodic & rad\\
$\phi_1$ & first-stage initial pitch & $[\pi/4,\ \pi/2]$ & rad\\
$\phi_2$ & second-stage initial pitch & $[-\pi/2,\ \pi/2]$ & rad\\
$\psi_2$ & second-stage yaw & $[-\pi,\ \pi)$, periodic & rad\\
$\dot\phi_2$ & second-stage pitch rate & $[-\pi/60,\ \pi/60]$ & rad/s\\
$t_c$ & coast duration & $[250,\ 400]$ & s\\
$T_2$ & second-stage burn time & $[0,\ 232.43]$ & s\\
$\alpha_m$ & 1st-stage max.\ angle of attack & $[-\pi/12,\ \pi/12]$ & rad\\
\hline
\end{tabular}
\end{table}

\paragraph{Terminal conditions} Let $R_{\mathrm{ref}}$ be the equatorial radius of Mars and $\mu$ its gravitational parameter. Let $\mathbf h_M$ be the angular momentum $\mathbf r_f\times\mathbf v_f$ expressed in the Mars-fixed frame, with $z$-component $h_{M,z}$. A circular orbit with radius $a^\star$ and inclination $i^\star$ requires the four terminal residuals
\begin{equation}
\mathbf F=
\begin{bmatrix}
\left(\|\mathbf r_f\|-a^\star\right)/R_{\mathrm{ref}}\\[2pt]
\|\mathbf r_f\|\,\|\mathbf v_f\|^2/\mu-1\\[2pt]
10\,\mathbf r_f^T\mathbf v_f/\left(\|\mathbf r_f\|\,\|\mathbf v_f\|\right)\\[2pt]
10\left(h_{M,z}/\|\mathbf h_M\|-\cos i^\star\right)
\end{bmatrix}
\label{eq:residual}
\end{equation}
to vanish. The four rows express the target radius, the circular speed, a zero radial velocity and the target inclination. The second and third rows together impose circularity; they avoid using the eccentricity itself, which is poorly conditioned near zero. The refinement in Section~\ref{sec:method} uses the weighted residual
\begin{equation}
\mathbf r(\mathbf z)=\mathbf W\,\mathbf F(\mathbf z),\qquad \mathbf W=\mathrm{diag}(100,1,1,1),
\end{equation}
where the dependence on $\boldsymbol\theta$ is omitted for brevity.

The residual $\mathbf F$ is used for computation only. The accuracy of a solution is judged on the orbital elements. A solution is accurate if
\begin{equation}
|a-a^\star|\le30~\mathrm m,\quad |e-e^\star|\le10^{-6},\quad |i-i^\star|\le10^{-6}~\mathrm{rad},
\label{eq:accuracy}
\end{equation}
where $a$, $e$ and $i$ are computed from the terminal state. The eccentricity is evaluated from the signed residuals $F_2$ and $F_3$ to avoid cancellation near $e=0$.

\paragraph{Dynamic-pressure constraint} The dynamic pressure at first-stage burnout, $q_{b1}=\tfrac12\rho\|\mathbf v-\boldsymbol\omega\times\mathbf r\|^2$, where $\boldsymbol\omega$ is the rotation rate of Mars, is limited to protect the vehicle during stage separation. The limit is set by the engineers to $\bar q=400$~Pa. It is not a hard physical boundary, and a tolerance of $q_{tol}=15$~Pa is accepted in operation. We write the constraint as
\begin{equation}
g(\mathbf z)=q_{b1}(\mathbf z)-\bar q\le0 .
\end{equation}

\subsection{Reference optimization problem and onboard determination problem}\label{sec:problems}

The ascent parameters are conventionally obtained by minimizing the second-stage burn time, which minimizes the second-stage propellant consumption $\dot m_2T_2$:
\begin{equation}
\text{(P)}\qquad
\begin{aligned}
\min_{\mathbf z}\quad & f(\mathbf z)=T_2\\
\text{s.t.}\quad & \mathbf r(\mathbf z)=\mathbf 0,\\
& g(\mathbf z)\le0,\\
& -1\le z_j\le1,\quad j\in\mathcal N .
\end{aligned}
\label{eq:P}
\end{equation}
Because $T_2$ is itself the seventh ascent parameter, $T_2=c_{x,7}+s_{x,7}z_7$ with $s_{x,7}=116.21$~s, and the objective is linear in $\mathbf z$. Problem~(P) is solved offline by SQP to generate the training data, and its solutions serve as the reference $\mathbf z^\ast$ in the evaluation.

Onboard, solving (P) to optimality is not required. The onboard determination problem is to find, for the measured mission conditions $\boldsymbol\theta$, ascent parameters $\hat{\mathbf z}$ that satisfy
\begin{equation}
\text{(F)}\quad
\begin{aligned}
&\hat{\mathbf z}\ \text{meets the accuracy requirement}~\eqref{eq:accuracy},\\
&q_{b1}(\hat{\mathbf z})\le\bar q+q_{tol},\quad
-1\le\hat z_j\le1,\ j\in\mathcal N,
\end{aligned}
\label{eq:F}
\end{equation}
within a short computation time and with a second-stage propellant consumption close to that of the reference solution. Since $\dot m_2$ is fixed, the relative propellant difference equals
\begin{equation}
\frac{\Delta m_{p2}}{m_{p2}^\ast}=\frac{\hat T_2-T_2^\ast}{T_2^\ast}.
\label{eq:fuel}
\end{equation}
We report this quantity for every solution. Problem~(F) is a feasibility problem: it has four equations and eight unknowns, so its solutions form a four-dimensional set, most of which consumes more propellant than $\mathbf z^\ast$. The method in Section~\ref{sec:method} solves (F) from a starting point close to $\mathbf z^\ast$, so that the solution found is also close to $\mathbf z^\ast$.

\subsection{Computational model: simulations and sequential rounds}\label{sec:cost}

The computation time of every method considered here is dominated by trajectory simulations. One simulation takes about 24~ms on a single core of an Intel Xeon Gold 6330 processor with the C++ implementation used in Section~\ref{sec:experiments}, whereas one forward pass of the neural network takes less than 1~ms. Linear-algebra operations on matrices of size $4\times8$ are negligible. We therefore measure the cost of a method by two quantities.

\begin{itemize}
\item The number of simulations $N_{\mathrm{sim}}$, which determines the computation time on a single core.
\item The number of sequential rounds $N_{\mathrm{rnd}}$. A round is a set of simulations whose inputs are all known before any of them is started. Simulations within a round can run on different cores, whereas a round can begin only after the previous round has returned.
\end{itemize}

If round $k$ contains $n_k$ simulations and $P$ cores are available, the computation time is approximately
\begin{equation}
t\approx t_{\mathrm{sim}}\sum_{k=1}^{N_{\mathrm{rnd}}}\left\lceil\frac{n_k}{P}\right\rceil ,
\label{eq:time}
\end{equation}
where $t_{\mathrm{sim}}$ is the mean time of one simulation. Because the fixed-step RK4 scheme integrates over the total flight duration $T_{b1}+t_c+T_2$, individual simulation runtimes scale approximately with flight time and vary modestly across missions and iterates; however, within each finite-difference Jacobian round the perturbation step ($h=10^{-5}$) is tiny, so all concurrent trajectories have virtually identical durations and incur negligible synchronization imbalance across cores. This justifies treating $t_{\mathrm{sim}}$ as uniform in~\eqref{eq:time} and using $N_{\mathrm{sim}}$ and $N_{\mathrm{rnd}}$ as hardware-independent cost metrics (alongside directly measured server wall-clock times in Section~\ref{sec:speed}). With one core, $t\approx N_{\mathrm{sim}}t_{\mathrm{sim}}$; with $P\ge\max_k n_k$ cores, $t\approx N_{\mathrm{rnd}}t_{\mathrm{sim}}$. On a multi-core processor whose single core is slow, a method is fast only if it has both few simulations and few rounds.

For SQP with finite-difference gradients, the gradient of one iterate requires $2\times8=16$ simulations that form one round. The line search, however, evaluates its trial points one after another, and each trial point forms a round of its own. Hence $N_{\mathrm{rnd}}$ of SQP equals the number of line-search evaluations plus the number of gradient evaluations, and additional cores reduce only the latter part.

\section{Learned Initial Guess and Parallel Boundary-Frozen Refinement}\label{sec:method}

\subsection{Overview}\label{sec:overview}

The framework decouples computation into an offline learning phase and an online feasibility-restoration phase (Fig.~\ref{fig:architecture}(a)). Offline, Problem~(P) is solved via SQP over sampled missions to train a ResNet mapping $\boldsymbol\theta$ to near-optimal ascent parameters. Online, the network predicts $\mathbf z_0$ in a single forward pass, and boundary-frozen damped SVD iteratively eliminates terminal orbital residuals without evaluating $T_2$, executing each iteration in two parallel simulation rounds across processor cores.

\subsection{Offline SQP database and ResNet initial guess}\label{sec:resnet}

\paragraph{SQP database generation} We sampled 30,000 missions across the operational domain defined in Section~\ref{sec:params} via Latin hypercube sampling over $(B_0,H_0,\kappa,i^\star)$ and solved Problem~(P) for each instance using SciPy's SLSQP solver \cite{virtanen2020scipy} coupled with the 0.025~s event-aligned integrator. Gradients of the terminal residuals and burnout dynamic pressure were evaluated via central differences with a normalized step size of $10^{-5}$ (switching to one-sided differences at parameter bounds), using a convergence tolerance of $10^{-9}$ and a budget of 30 iterations. Initialized from a fixed engineering guess with $A_0=\pi/2+i^\star$, 73\% of the runs converged within the strict $10^{-9}$ tolerance, and all trajectories remained within the southeast launch family $A_0\in(\pi/2,\pi)$. The dataset was randomly partitioned into 24,000 training, 1,500 validation, 1,500 calibration, and 3,000 test missions. For each mission, the training label $\mathbf z_{\mathrm{label}}$ was selected as the results satisfying the terminal tolerances alongside the dynamic-pressure limit. Thus, every training label lies in the immediate vicinity of the true optimum of Problem~(P).

\paragraph{Network architecture and training} The predictor $\mathcal R_\phi$ maps the normalized mission conditions $\tilde{\boldsymbol\theta}\in\mathbb R^6$ to the normalized initial guess:
\begin{equation}
\mathbf z_0=\Pi\left(\mathcal R_\phi(\tilde{\boldsymbol\theta})\right),
\end{equation}
where each component of $\boldsymbol\theta$ is centered by its training-set median and scaled by the 90th percentile of its absolute deviation, and the projection operator $\Pi(\cdot)$ enforces domain feasibility by clipping bounded coordinates to $[-1,1]$ and wrapping periodic angles. The network comprises an input linear projection from 6 to 128 units with GELU activation \cite{hendrycks2016gelu}, four residual blocks \cite{he2016resnet} of width 128, and a linear output layer mapping 128 to 8 units. Each residual block evaluates
\begin{equation}
\mathbf h^{+}=\mathrm{GELU}\!\left(\mathbf h+\mathbf B_2\,\mathrm{GELU}\!\left(\mathbf B_1\,\mathrm{LN}(\mathbf h)+\mathbf b_1\right)+\mathbf b_2\right),
\end{equation}
where $\mathrm{LN}(\cdot)$ denotes layer normalization \cite{ba2016layernorm}. The model contains approximately $1.35\times10^5$ trainable parameters, requiring only 0.5~MB of memory in 32-bit single precision. We trained the network using the smooth-$L_1$ (Huber) loss on normalized coordinates (with transition threshold 0.1) via the AdamW optimizer (initial learning rate $3\times10^{-4}$, weight decay $10^{-4}$, batch size 256, and gradient norm clipping at 5.0) under a cosine annealing schedule over 180 epochs. Early stopping with a 25-epoch patience selected the best checkpoint at epoch 165.

Across the validation set, the prediction error $\|\mathbf z_0-\mathbf z_{\mathrm{label}}\|_2$ achieves a median of 0.0044 and a 95th percentile of 0.017 in normalized units. While this proximity is well within the local convergence basin of our refinement solver, it is insufficient for direct orbit insertion: zero out of the 3,000 test predictions satisfy the terminal accuracy tolerances in~\eqref{eq:accuracy}.

\subsection{Boundary-frozen damped SVD refinement}\label{sec:svd}

Starting from $\mathbf z_0$, the online solver drives the weighted terminal residual $\mathbf r(\mathbf z)$ to zero via the iterative procedure summarized in Algorithm~\ref{alg:svd}.

\paragraph{Boundary freezing} Any box-constrained coordinate $j\in\mathcal N$ lying within a margin $\beta_f=0.002$ (0.1\% of its full range) of its upper or lower bound is frozen at its current value and excluded from subsequent updates. An initial frozen set $\mathcal F_0$ is identified at $\mathbf z_0$ and remains locked throughout the solve, while additional coordinates reaching the $\beta_f$-margin at later iterations are frozen dynamically. Without boundary freezing, update steps crossing a bound would be truncated by $\Pi(\cdot)$, distorting the search direction and potentially inducing stagnation; freezing also obviates one-sided finite differencing at the boundary. The remaining active coordinates form the free index set $\mathcal I$, with cardinality $m=|\mathcal I|$. Across all 3,000 test missions, the ResNet initialization leaves at least $m\ge6$ free coordinates.

\paragraph{Jacobian evaluation} For each free coordinate $j\in\mathcal I$, the corresponding Jacobian column is approximated via central finite differences with a normalized perturbation step $h=10^{-5}$:
\begin{equation}
\mathbf J_{:,j}=\frac{\mathbf r(\mathbf z+h\mathbf e_j)-\mathbf r(\mathbf z-h\mathbf e_j)}{2h},
\end{equation}
requiring $2m\le16$ mutually independent trajectory simulations, which execute in a single parallel batch. The residual $\mathbf r(\mathbf z)$ at the current iterate is also needed for the step. For $k\ge2$ it is available from the accepted candidate of the preceding iteration. At the first iteration, the nominal trajectory at $\mathbf z_0$ is not simulated; instead, $\mathbf r(\mathbf z_0)$ and the dynamic pressure are estimated by the mean over the perturbation pairs, $\bar{\mathbf r}=\frac1{2m}\sum_{j\in\mathcal I}\left[\mathbf r(\mathbf z_0+h\mathbf e_j)+\mathbf r(\mathbf z_0-h\mathbf e_j)\right]$, whose error is $O(h^2)$. On all test missions, this estimate gives the same number of iterations and the same success as a simulation of $\mathbf z_0$, while keeping the first batch at 16 simulations, which fits a processor with 8 or 16 cores. (If every coordinate were frozen, $\mathbf z_0$ would be simulated instead.)

\paragraph{Damped minimum-norm step} Let $\mathbf J_{\mathcal I}=\mathbf U\boldsymbol\Sigma\mathbf V^T$ denote the thin singular value decomposition of the reduced Jacobian $\mathbf J_{\mathcal I}\in\mathbb R^{4\times m}$, where $\mathbf U\in\mathbb R^{4\times 4}$, $\mathbf V\in\mathbb R^{m\times 4}$, and $\boldsymbol\Sigma=\mathrm{diag}(\sigma_1,\sigma_2,\sigma_3,\sigma_4)$ with singular values arranged in descending order $\sigma_1\ge\sigma_2\ge\sigma_3\ge\sigma_4\ge 0$. The update vector on the free coordinates solves the Levenberg--Marquardt damped least-squares problem $\min_{\mathbf v}\|\mathbf r+\mathbf J_{\mathcal I}\mathbf v\|_2^2+\mu^2\|\mathbf v\|_2^2$ \cite{levenberg1944,marquardt1963}:
\begin{equation}
\mathbf v_\mu=-\mathbf V\,\mathrm{diag}\!\left(\frac{\sigma_i}{\sigma_i^2+\mu^2}\right)\mathbf U^T\mathbf r,\qquad
\mu=\gamma\,\sigma_1,
\label{eq:svdstep}
\end{equation}
initialized with a relative damping factor $\gamma=10^{-4}$. For small $\gamma$, $\mathbf v_\mu$ closely approximates the Moore--Penrose pseudoinverse step $-\mathbf J_{\mathcal I}^{+}\mathbf r$, which eliminates the terminal residual while remaining orthogonal to the null space $\ker(\mathbf J_{\mathcal I})$. Consequently, the update suppresses first-order drift along the feasible manifold $\mathbf r(\mathbf z)=\mathbf 0$, preserving the propellant near-optimality inherited from $\mathbf z_0$ (Section~\ref{sec:theory}). The full-space direction $\mathbf d\in\mathbb R^8$ is assembled by mapping $\mathbf v_\mu$ into the free indices $\mathcal I$ and setting frozen indices to zero, followed by trust-region norm scaling such that $\|\mathbf d\|_2\le\rho$ (initialized at $\rho=0.05$).

\paragraph{Parallel trial steps and acceptance} To eliminate serial line searches, four candidate iterates
\begin{equation}
\mathbf z_\alpha=\Pi(\mathbf z+\alpha\mathbf d),\qquad \alpha\in\{1,\tfrac12,\tfrac14,\tfrac18\},
\end{equation}
are evaluated concurrently in a single parallel round. A candidate $\mathbf z_\alpha$ is deemed \emph{admissible} if its trajectory integration succeeds, lies within parameter bounds, and satisfies the relaxed dynamic-pressure limit $q_{b1}(\mathbf z_\alpha)\le\bar q+q_{tol}$. Candidates are inspected in descending order of $\alpha$: if any admissible $\mathbf z_\alpha$ satisfies the orbital insertion tolerances in~\eqref{eq:accuracy}, the first such point is returned and the algorithm terminates. Otherwise, the first admissible candidate achieving a strict residual reduction $\|\mathbf r(\mathbf z_\alpha)\|_2<\|\mathbf r(\mathbf z)\|_2$ is accepted as the next iterate, and both $\gamma$ and $\rho$ are reset to their nominal values. If no candidate reduces the residual, the iterate $\mathbf z$ is retained, damping is increased ($\gamma\gets10\gamma$), and the trust radius is contracted ($\rho\gets\rho/2$). 

The primary loop is capped at 15 iterations. As a safeguard against rare non-convergence, the flight software is configured to reset $(\gamma,\rho)$ and continue for an extended budget of 100 iterations; if still unsuccessful, onboard generation aborts and awaits ground-uplinked parameters. Across all 3,000 test missions, neither fallback stage was ever triggered.

\begin{algorithm}[t]
\caption{Boundary-frozen damped SVD refinement}\label{alg:svd}
\begin{algorithmic}[1]
\Require Mission condition $\boldsymbol\theta$, network prediction $\mathbf z_0$
\State $\mathbf z\gets\mathbf z_0$, $\gamma\gets10^{-4}$, $\rho\gets0.05$, $\mathcal F_0\gets\{j\in\mathcal N : |z_{0,j}|\ge 1-\beta_f\}$
\For{$k=1,\dots,15$}
  \State $\mathcal I\gets\{1,\dots,8\}\setminus\left(\mathcal F_0\cup\{j\in\mathcal N : |z_j|\ge 1-\beta_f\}\right)$
  \State \textbf{Round A:} Simulate $\mathbf z\pm h\mathbf e_j$ for all $j\in\mathcal I$ in parallel
  \State Assemble $\mathbf J_{\mathcal I}$; if $k=1$, set $\mathbf r(\mathbf z)\gets\bar{\mathbf r}$ (mean over the pairs); compute $\mathbf d$ via~\eqref{eq:svdstep} and scale if $\|\mathbf d\|_2>\rho$
  \State \textbf{Round B:} Simulate $\mathbf z_\alpha=\Pi(\mathbf z+\alpha\mathbf d)$ for $\alpha\in\{1,\frac12,\frac14,\frac18\}$ in parallel
  \If{any admissible $\mathbf z_\alpha$ satisfies terminal tolerances~\eqref{eq:accuracy}}
    \State \Return first such $\mathbf z_\alpha$ (ordered by descending $\alpha$)
  \ElsIf{any admissible $\mathbf z_\alpha$ satisfies $\|\mathbf r(\mathbf z_\alpha)\|_2<\|\mathbf r(\mathbf z)\|_2$}
    \State $\mathbf z\gets$ first such $\mathbf z_\alpha$; $\gamma\gets10^{-4}$; $\rho\gets0.05$
  \Else
    \State $\gamma\gets10\gamma$; $\rho\gets\rho/2$
  \EndIf
\EndFor
\State \Return failure
\end{algorithmic}
\end{algorithm}

\subsection{Parallel batch structure and cost}\label{sec:parallel}

\begin{figure*}[t]
\centering
\includegraphics[width=\textwidth]{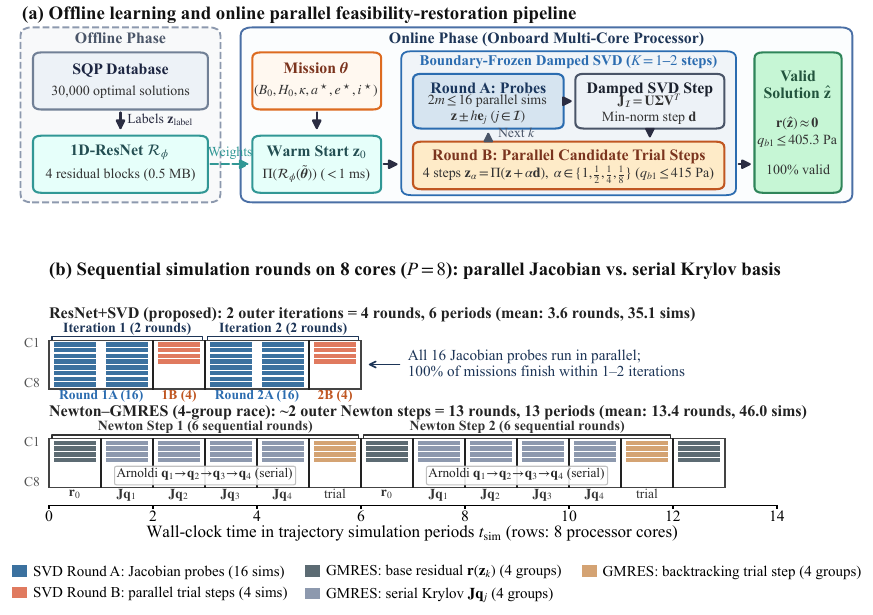}
\caption{Proposed onboard ascent parameter determination framework and parallel round structure. (a)~Offline SQP database generation and 1D-ResNet training paired with online boundary-frozen damped SVD refinement. (b)~Core-level concurrency across sequential simulation rounds on an eight-core processor ($P=8$), contrasting a two-iteration solve of ResNet+SVD (four rounds, six simulation periods $t_{\mathrm{sim}}$) against the four-group Newton--GMRES race, in which each outer Newton step requires six sequential rounds due to serial Arnoldi/Krylov basis construction ($\mathbf q_1\to\mathbf q_2\to\mathbf q_3\to\mathbf q_4$).}\label{fig:architecture}
\end{figure*}

Each iteration of Algorithm~\ref{alg:svd} executes in two sequential rounds (Fig.~\ref{fig:architecture}(b)): \textbf{Round~A} evaluates the $2m\le16$ central-difference perturbations (with $\mathbf r(\mathbf z_0)$ estimated by $\bar{\mathbf r}$ at $k=1$ and reused thereafter), and \textbf{Round~B} evaluates the four trial steps concurrently. A $K$-iteration solve therefore incurs
\begin{equation}
N_{\mathrm{sim}}=\sum_{k=1}^{K}\left(2m_k+4\right),\qquad N_{\mathrm{rnd}}=2K,
\end{equation}
where $m_k=|\mathcal I_k|$. Across the 3,000 test missions, refinement converges in $K=1$ iteration (35\%) or $K=2$ iterations (65\%), averaging $N_{\mathrm{sim}}=32$ simulations and $N_{\mathrm{rnd}}=3.3$ rounds.

From~\eqref{eq:time}, the runtime on $P\ge4$ cores is approximated by
\begin{equation}
t\approx t_{\mathrm{sim}}\sum_{k=1}^{K}\left(\left\lceil\frac{n_k}{P}\right\rceil+1\right),
\end{equation}
where $n_k=2m_k\le16$ and Round~B takes one period $t_{\mathrm{sim}}$. Thus, Round~A takes two periods on 8 cores and one period when $P\ge16$ (compressing total runtime to $\approx 3.3\,t_{\mathrm{sim}}$), whereas Round~B evaluates all four trial steps simultaneously in one period instead of running serial backtracking trials.

\section{Local Analysis: Convergence and Propellant Penalty}\label{sec:theory}

This section addresses two fundamental theoretical questions: (i) how accurate must the learned prediction be to guarantee convergence of the online refinement to a feasible solution of Problem~(F), and (ii) what propellant penalty does the resulting trajectory incur relative to the optimal reference solution of Problem~(P)? Our analysis is local, holding within a neighborhood of a reference solution $\mathbf z^\ast$ to establish sufficient convergence conditions and asymptotic error orders rather than tight global constants. Detailed proofs omitted from the main text are provided in Appendix~\ref{app:proofs}.

Throughout this section, let $\mathbf z_0=\mathbf z^\ast+\mathbf e$ denote the network prediction with initial error $\mathbf e$, and let $\hat{\mathbf z}$ denote the refined output with final deviation $\boldsymbol\delta=\hat{\mathbf z}-\mathbf z^\ast$ and Euclidean norm $D=\|\boldsymbol\delta\|$. We denote the initial and final residual norms by $R_0=\|\mathbf r(\mathbf z_0)\|$ and $\varepsilon_c=\|\mathbf r(\hat{\mathbf z})\|$, respectively. Unless stated otherwise, $\|\cdot\|$ denotes the Euclidean norm for vectors and the spectral norm for matrices.

\subsection{Convergence under a bounded initial-guess error}\label{sec:convergence}

Let $\mathcal U$ be a convex neighborhood of $\mathbf z^\ast$, and let $\mathcal B\subset\mathcal U$ denote the subset of points that coincide with $\mathbf z_0$ on the frozen coordinates and lie within a radius $r_B$ of $\mathbf z_0$ on the free coordinates. We denote by $\mathbf J_{\mathcal I}(\mathbf z)$ the reduced Jacobian of $\mathbf r$ with respect to the free index set $\mathcal I$, by $\tilde{\mathbf J}_k$ its central-difference approximation at iteration $k$, and by $\mu_k=\gamma\sigma_1(\tilde{\mathbf J}_k)$ the damping parameter in~\eqref{eq:svdstep}. We impose the following local assumptions:
\begin{description}
\item[(A1) Smoothness.] The sequence of flight phases remains invariant across $\mathcal U$. Thus, $\mathbf r$ is continuously differentiable on $\mathcal U$, and its full Jacobian $\mathbf J$ is Lipschitz continuous with constant $L$ and bounded above by $M_J=\sup_{\mathbf z\in\mathcal U}\|\mathbf J(\mathbf z)\|$.
\item[(A2) Full row rank.] For every admissible free index set $\mathcal I$, the minimum singular value satisfies $\sigma_{\min}(\mathbf J_{\mathcal I}(\mathbf z))\ge\sigma>0$ for all $\mathbf z\in\mathcal B$. Physically, the free ascent parameters retain independent authority over all four terminal orbital conditions.
\item[(A3) Finite-difference accuracy.] The numerical Jacobian error satisfies $\|\tilde{\mathbf J}_k-\mathbf J_{\mathcal I}(\mathbf z_k)\|\le\varepsilon_J<\sigma$, defining the effective singular value lower bound $\tilde\sigma=\sigma-\varepsilon_J>0$ (empirically, our benchmark exhibits a median $\sigma_{\min}(\tilde{\mathbf J}_k)\approx 0.80$, while the step-sensitivity verification in Section~\ref{sec:model} confirms negligible differencing noise at $h=10^{-5}$).
\item[(A4) Small damping.] With $\bar\mu=\sup_k\mu_k$, the composite perturbation factor $\chi=(\bar\mu/\tilde\sigma)^2+\varepsilon_J/\tilde\sigma$ satisfies $\chi\le1/4$. For the nominal damping $\gamma=10^{-4}$ and negligible differencing error $\varepsilon_J$, this holds whenever the singular-value ratio $\sigma_1(\tilde{\mathbf J}_k)/\tilde\sigma$ (an upper bound on the condition number of $\tilde{\mathbf J}_k$) remains below 5,000 (around 37 in our benchmark).
\item[(A5) Local admissibility and invariant bounds.] Within $\mathcal B$, all free coordinates remain strictly farther than $\beta_f$ from their box bounds (preventing boundary clipping and additional coordinate freezing), and every simulated trajectory integrates validly with $q_{b1}\le\bar q+q_{tol}$. By the descending inspection order of $\alpha$ in Algorithm~\ref{alg:svd}, whenever the full step ($\alpha=1$) is admissible and reduces $\|\mathbf r\|$, the accepted candidate yields a residual norm no larger than $\|\mathbf r(\mathbf z_k+\mathbf d_k)\|$.
\end{description}
Assumption~(A1) holds because event-aligned integration preserves smoothness whenever the switching sequence is unchanged, while (A5) formalizes the local regime of the boundary-freezing, dynamic-pressure, and parallel acceptance rules in Algorithm~\ref{alg:svd}. Missions violating these local conditions are evaluated empirically in Section~\ref{sec:experiments}.

\begin{lemma}[One-step contraction]\label{lem:contraction}
Under Assumptions~(A1)--(A4), if $\mathbf z_k\in\mathcal B$ and $\mathbf z_k+\mathbf d_k\in\mathcal B$, the damped SVD step $\mathbf d_k$ satisfies (with $\mathbf r_k=\mathbf r(\mathbf z_k)$)
\begin{equation}
\|\mathbf d_k\|\le\frac{\|\mathbf r_k\|}{\tilde\sigma},\qquad
\|\mathbf r(\mathbf z_k+\mathbf d_k)\|\le\left(\chi+\frac{L\|\mathbf r_k\|}{2\tilde\sigma^2}\right)\|\mathbf r_k\| .
\end{equation}
\end{lemma}

\begin{theorem}[Local convergence to a feasible point]\label{thm:convergence}
Under Assumptions~(A1)--(A5), if the initial residual satisfies
\begin{equation}
R_0\le\frac{\tilde\sigma^2}{2L},\qquad \frac{2R_0}{\tilde\sigma}\le r_B,\qquad \frac{R_0}{\tilde\sigma}\le\rho_0,
\label{eq:thm1cond}
\end{equation}
where $\rho_0=0.05$ is the initial trust-region radius, then the iterates satisfy $\|\mathbf r_k\|\le2^{-k}R_0$ and $\|\mathbf z_k-\mathbf z_0\|\le2R_0/\tilde\sigma$ for all $k\ge0$. Consequently, the residual norm reaches any target tolerance $\tau>0$ within at most $\lceil\log_2(R_0/\tau)\rceil$ iterations, and in the absence of early termination, $\mathbf z_k$ converges to a feasible limit $\mathbf z_\infty\in\mathcal B$ satisfying $\mathbf r(\mathbf z_\infty)=\mathbf 0$. Furthermore, with $\zeta_k=L\|\mathbf r_k\|/(2\tilde\sigma^2)$, each step contracts quadratically ($\zeta_{k+1}\le2\zeta_k^2$) while $\zeta_k\ge\chi$, and linearly ($\zeta_{k+1}\le2\chi\zeta_k$) once $\zeta_k<\chi$.
\end{theorem}

\begin{corollary}[Admissible network prediction error]\label{cor:emax}
Suppose Assumptions~(A1)--(A5) hold uniformly for all initial predictions satisfying $\|\mathbf z_0-\mathbf z^\ast\|\le\bar e$, with the line segment connecting $\mathbf z^\ast$ and $\mathbf z_0$ contained in $\mathcal U$. Since $\mathbf r(\mathbf z^\ast)=\mathbf 0$, the mean-value bound yields $R_0\le M_J\|\mathbf e\|$. Thus, whenever
\begin{equation}
\|\mathbf e\|\le e_{\max}=\min\left\{\bar e,\ \frac{\tilde\sigma^2}{2LM_J},\ \frac{\tilde\sigma r_B}{2M_J},\ \frac{\tilde\sigma\rho_0}{M_J}\right\},
\label{eq:emax}
\end{equation}
Algorithm~\ref{alg:svd} converges from $\mathbf z_0$ to a feasible solution $\hat{\mathbf z}$ of Problem~(F) satisfying
\begin{equation}
\begin{aligned}
\|\hat{\mathbf z}-\mathbf z_0\|&\le\frac{2M_J}{\tilde\sigma}\|\mathbf e\|,\\
D=\|\hat{\mathbf z}-\mathbf z^\ast\|&\le c_e\|\mathbf e\|,\qquad c_e=1+\frac{2M_J}{\tilde\sigma}.
\end{aligned}
\label{eq:Dbound}
\end{equation}
\end{corollary}

Corollary~\ref{cor:emax} formalizes the exact role of the neural network: placing $\mathbf z_0$ within the convergence radius $e_{\max}$. While a cold start generally violates~\eqref{eq:emax} and lacks theoretical guarantees, an accurate ResNet prediction initializes the solver with $\zeta_0\ll1/4$; although Theorem~\ref{thm:convergence} guarantees asymptotic convergence within $\lceil\log_2(R_0/\tau)\rceil$ steps, the actual per-step contraction factor $\chi+\zeta_k$ in Lemma~\ref{lem:contraction} is orders of magnitude smaller than the worst-case bound of $1/2$, explaining why $K\le2$ iterations suffice empirically across all test missions (well within the 15-iteration budget). Notably, frozen coordinates require no correction: the bound $R_0\le M_J\|\mathbf e\|$ accounts for errors across all eight coordinates, whereas the iterative updates act exclusively on the well-conditioned free subspace.

\subsection{Propellant penalty inherited from the initial guess}\label{sec:penalty}

Because the refinement does not minimize $T_2$, its output $\hat{\mathbf z}$ deviates slightly from $\mathbf z^\ast$. We now characterize the burn-time excess $\Delta T_2=f(\hat{\mathbf z})-f(\mathbf z^\ast)$, which is directly proportional to the relative propellant penalty via~\eqref{eq:fuel}. Our analysis encompasses boundary-active solutions: across our benchmark, roughly one-third of the reference optima ride the dynamic-pressure limit $g(\mathbf z^\ast)=0$, about 14\% saturate the lower bound of $\phi_1$, and about 21\% the upper bound of the coast duration $t_c$.

\paragraph{First-order optimality conditions} At the reference optimum $\mathbf z^\ast$, let the active box-bound sets be $\mathcal A^{up}=\{j\in\mathcal N:z^\ast_j=1\}$ and $\mathcal A^{lo}=\{j\in\mathcal N:z^\ast_j=-1\}$, with $\mathcal A_B=\mathcal A^{up}\cup\mathcal A^{lo}$. We assume standard regularity at $\mathbf z^\ast$:
\begin{description}
\item[(A6) Regularity (LICQ and strict complementarity).] The gradients of all active constraints at $\mathbf z^\ast$---namely, the rows of $\mathbf J(\mathbf z^\ast)$, $\nabla g(\mathbf z^\ast)^T$ (if $g(\mathbf z^\ast)=0$), and $\mathbf e_j^T$ for $j\in\mathcal A_B$---are linearly independent, and the Lagrange multipliers associated with active inequality constraints are strictly positive.
\end{description}
Under~(A6), there exist unique Karush--Kuhn--Tucker (KKT) multipliers $\boldsymbol\lambda\in\mathbb R^4$, $\nu\ge0$, and $\eta_j>0$ ($j\in\mathcal A_B$) satisfying
\begin{equation}
\begin{aligned}
\nabla f+\mathbf J(\mathbf z^\ast)^T\boldsymbol\lambda+\nu\nabla g(\mathbf z^\ast)\qquad\qquad&\\
+\sum_{j\in\mathcal A^{up}}\eta_j\mathbf e_j-\sum_{j\in\mathcal A^{lo}}\eta_j\mathbf e_j&=\mathbf 0,\\
\nu\,g(\mathbf z^\ast)&=0 .
\end{aligned}
\label{eq:kkt}
\end{equation}
Because $\nabla f=s_{x,7}\mathbf e_7$ is constant, these multipliers can be recovered directly from the active constraint gradients at $\mathbf z^\ast$ via least squares. For each $j\in\mathcal A_B$, let $b_j(\hat{\mathbf z})\ge0$ denote the inward distance from $\hat z_j$ to its active bound ($1-\hat z_j$ for $j\in\mathcal A^{up}$ and $\hat z_j+1$ for $j\in\mathcal A^{lo}$), which is non-negative because $\hat{\mathbf z}$ respects box bounds. Finally, let $L_g$ be the Lipschitz constant of $\nabla g$ on $\mathcal U$ and $G_q=\sup_{\mathbf z\in\mathcal U}\|\nabla g(\mathbf z)\|$, and define the second-order Taylor remainders
\begin{equation}
\mathbf R_r=\mathbf r(\hat{\mathbf z})-\mathbf J(\mathbf z^\ast)\boldsymbol\delta,\quad
R_g=g(\hat{\mathbf z})-g(\mathbf z^\ast)-\nabla g(\mathbf z^\ast)^T\boldsymbol\delta,
\end{equation}
which obey $\|\mathbf R_r\|\le\tfrac12 LD^2$ and $|R_g|\le\tfrac12 L_gD^2$.

\begin{theorem}[Exact decomposition of the propellant penalty]\label{thm:penalty}
Under Assumptions~(A1) and~(A6), any refined solution $\hat{\mathbf z}$ satisfies the exact identity
\begin{equation}
\Delta T_2=
\underbrace{-\boldsymbol\lambda^T\mathbf r(\hat{\mathbf z})}_{T_c}
\underbrace{-\nu\,g(\hat{\mathbf z})}_{T_q}
+\underbrace{\sum_{j\in\mathcal A_B}\eta_j\,b_j(\hat{\mathbf z})}_{T_B}
+\underbrace{\boldsymbol\lambda^T\mathbf R_r+\nu R_g}_{T^{(2)}} .
\label{eq:penalty}
\end{equation}
\end{theorem}

\begin{proof}
By linearity of $f$, $\Delta T_2=\nabla f^T\boldsymbol\delta$. Substituting $\nabla f$ from the KKT stationarity condition~\eqref{eq:kkt} yields
\begin{equation*}
\begin{aligned}
\Delta T_2=&-\boldsymbol\lambda^T\mathbf J(\mathbf z^\ast)\boldsymbol\delta-\nu\nabla g(\mathbf z^\ast)^T\boldsymbol\delta\\
&-\sum_{j\in\mathcal A^{up}}\eta_j\delta_j+\sum_{j\in\mathcal A^{lo}}\eta_j\delta_j .
\end{aligned}
\end{equation*}
Since $\mathbf r(\mathbf z^\ast)=\mathbf 0$, we have $-\boldsymbol\lambda^T\mathbf J(\mathbf z^\ast)\boldsymbol\delta=-\boldsymbol\lambda^T(\mathbf r(\hat{\mathbf z})-\mathbf R_r)$. Similarly, complementary slackness $\nu g(\mathbf z^\ast)=0$ implies $-\nu\nabla g(\mathbf z^\ast)^T\boldsymbol\delta=-\nu(g(\hat{\mathbf z})-R_g)$. For $j\in\mathcal A^{up}$, $z^\ast_j=1$ gives $-\delta_j=1-\hat z_j=b_j(\hat{\mathbf z})$; for $j\in\mathcal A^{lo}$, $z^\ast_j=-1$ gives $\delta_j=\hat z_j+1=b_j(\hat{\mathbf z})$. Because box bounds are linear, they introduce no higher-order remainder.
\end{proof}

Identity~\eqref{eq:penalty} holds exactly without requiring $\hat{\mathbf z}$ to be optimal or even strictly feasible with respect to $\bar q$. Its four constituent terms admit clear physical interpretations:
\begin{itemize}
\item $T_c$ is the \emph{terminal-residual term}, bounded by $|T_c|\le\|\boldsymbol\lambda\|\varepsilon_c$, which is rendered negligible by the tight orbital insertion tolerances.
\item $T_q=\nu(\bar q-q_{b1}(\hat{\mathbf z}))$ is the \emph{dynamic-pressure drift term}, which vanishes whenever $g(\mathbf z^\ast)<0$ ($\nu=0$). When $g(\mathbf z^\ast)=0$, drifting strictly inside the limit ($q_{b1}(\hat{\mathbf z})<\bar q$) incurs a positive propellant penalty ($T_q>0$), whereas slight excursions into the operational buffer ($\bar q<q_{b1}(\hat{\mathbf z})\le\bar q+q_{tol}$) yield a small negative term bounded below by $T_q\ge-\nu q_{tol}$.
\item $T_B\ge0$ is the \emph{bound-detachment term}, which vanishes if no parameter bound is active at $\mathbf z^\ast$ or if $\hat{\mathbf z}$ remains locked on the active bound.
\item $T^{(2)}$ is the \emph{second-order curvature term}, bounded by $|T^{(2)}|\le\tfrac12(\|\boldsymbol\lambda\|L+\nu L_g)D^2$.
\end{itemize}

\begin{corollary}[Asymptotic order of the propellant penalty]\label{cor:order}
Under the hypotheses of Corollary~\ref{cor:emax} and Assumption~(A6), the burn-time excess satisfies
\begin{equation}
\left|\Delta T_2-T_q-T_B\right|\le\|\boldsymbol\lambda\|\varepsilon_c+\tfrac12\left(\|\boldsymbol\lambda\|L+\nu L_g\right)c_e^2\|\mathbf e\|^2,
\end{equation}
with first-order inequality drift terms bounded by
\begin{equation}
\begin{aligned}
-\nu q_{tol}&\le T_q\le\nu\left(G_qc_e\|\mathbf e\|+\tfrac{L_g}{2}c_e^2\|\mathbf e\|^2\right),\\
0&\le T_B\le c_e\|\mathbf e\|\sum_{j\in\mathcal A_B}\eta_j .
\end{aligned}
\end{equation}
\end{corollary}

Corollary~\ref{cor:order} distinguishes three operational regimes:
\begin{enumerate}
\item \emph{Inactive inequalities at $\mathbf z^\ast$:} Here $T_q=T_B=0$, and $\Delta T_2=O(\varepsilon_c)+O(\|\mathbf e\|^2)$. The propellant penalty scales quadratically with the network prediction error.
\item \emph{Active inequalities preserved at $\hat{\mathbf z}$:} Here $T_q=T_B=0$, and the penalty again scales quadratically as $O(\|\mathbf e\|^2)$, because first-order objective variations vanish along the tangent manifold of all active constraints.
\item \emph{Active inequalities perturbed at $\hat{\mathbf z}$:} Detaching from or crossing an active boundary introduces a first-order term $O(\|\mathbf e\|)$ proportional to the corresponding multiplier ($\nu$ or $\eta_j$). Backing away from an active boundary incurs a propellant loss, whereas slightly exceeding $\bar q$ within the $q_{tol}$ buffer reduces $T_2$ below $T_2^\ast$ by at most $\nu q_{tol}$.
\end{enumerate}
Because the online refinement enforces only the four equality constraints $\mathbf r(\mathbf z)=\mathbf 0$ without tracking active inequalities, $q_{b1}(\hat{\mathbf z})$ and $b_j(\hat{\mathbf z})$ drift linearly with the initial prediction error $\mathbf e$. Consequently, keeping first-order penalties small relies directly on the accuracy of the learned warm start.

\paragraph{Role of boundary freezing} Across our dataset, active parameter bounds occur almost exclusively on $\phi_1$ and $t_c$ (and on $\alpha_m$ in only 8 of the 3,000 test missions). When the raw network prediction exceeds the bound, the projection $\Pi(\cdot)$ clamps it directly onto the boundary where it is immediately frozen, yielding $b_j(\hat{\mathbf z})=0$. If the prediction falls slightly inside the bound within the freezing margin $\beta_f$, it is frozen in place, capping the detachment at $b_j(\hat{\mathbf z})\le\beta_f=0.002$. A first-order bound penalty arises only when the network underpredicts an active bound by more than $\beta_f$. Boundary freezing thus strongly suppresses $T_B$ whenever the network places $\phi_1$ or $t_c$ near its active limit.

\paragraph{Row- and null-space error decomposition} A finer subspace analysis reveals exactly which component of the initial error $\mathbf e$ governs propellant suboptimality. Let $\mathbf P=\mathbf J^{\ast+}\mathbf J^\ast$, where $\mathbf J^\ast=\mathbf J(\mathbf z^\ast)$, denote the orthogonal projector onto the row space of $\mathbf J^\ast$. We decompose the initial error into orthogonal components $\mathbf e=\mathbf e_R+\mathbf e_N$, where $\mathbf e_R=\mathbf P\mathbf e\in\mathrm{row}(\mathbf J^\ast)$ drives the linearized terminal residual, and $\mathbf e_N=(\mathbf I-\mathbf P)\mathbf e\in\ker(\mathbf J^\ast)$ lies tangent to the feasible manifold $\mathbf r(\mathbf z)=\mathbf 0$.

\begin{proposition}[Minimum-norm correction and tangent drift]\label{prop:tangent}
Under the hypotheses of Corollary~\ref{cor:emax} and Assumption~(A6), suppose in addition that $\mathbf r$ and $g$ are twice continuously differentiable with Lipschitz continuous Hessians on $\mathcal U$ (constants $L_2$ and $L_{g,2}$, respectively), and consider the update on the free subspace (with no active freezing, where the unclipped full step $\alpha=1$ is accepted). Then the final deviation and dynamic-pressure drift satisfy
\begin{equation}
\begin{aligned}\boldsymbol\delta&=\mathbf e_N+O(\|\mathbf e\|^2)+O(\chi)\|\mathbf e\|,\\ g(\hat{\mathbf z})-g(\mathbf z^\ast)&=\nabla g(\mathbf z^\ast)^T\mathbf e_N+O(\|\mathbf e\|^2)+O(\chi)\|\mathbf e\|,\end{aligned}
\end{equation}
and the propellant penalty admits the second-order expansion
\begin{equation}
\begin{aligned}
\Delta T_2=&-\boldsymbol\lambda^T\mathbf r(\hat{\mathbf z})-\nu g(\hat{\mathbf z})+\sum_{j\in\mathcal A_B}\eta_j b_j(\hat{\mathbf z})\\
&+\tfrac12\mathbf e_N^T\nabla^2_{\mathbf z\mathbf z}\mathcal L(\mathbf z^\ast)\,\mathbf e_N+O\!\left(\|\mathbf e\|^3+\chi\|\mathbf e\|^2\right),
\end{aligned}
\label{eq:tangent}
\end{equation}
where $\mathcal L(\mathbf z)=f(\mathbf z)+\boldsymbol\lambda^T\mathbf r(\mathbf z)+\nu g(\mathbf z)$ is the Lagrangian. The constants hidden in the $O(\cdot)$ terms depend only on the constants of (A1)--(A5), on the Hessian-Lipschitz constants $L_2,L_{g,2}$ and on the multipliers, not on $\mathbf e$. If, in addition, $\mathbf z^\ast$ satisfies the second-order sufficient condition and $T_q=T_B=0$ at $\hat{\mathbf z}$, then $\tfrac12\mathbf e_N^T\nabla^2_{\mathbf z\mathbf z}\mathcal L(\mathbf z^\ast)\mathbf e_N\ge-O(\|\mathbf e\|^3+\chi\|\mathbf e\|^2)$.
\end{proposition}

Proposition~\ref{prop:tangent} clarifies the geometric role of the minimum-norm SVD step: because the linearized terminal residual $\mathbf J^\ast\mathbf e=\mathbf J^\ast\mathbf e_R$ contains no information about $\mathbf e_N$, the minimum-norm update is the unique step that annihilates the row-space error $\mathbf e_R$ to leading order without injecting an unobservable null-space perturbation, thereby leaving $\mathbf e_N$ undisturbed. When $\mathbf e_N$ lies in the tangent subspace of all active constraints ($T_q=T_B=0$), only the quadratic curvature term $\tfrac12\mathbf e_N^T\nabla^2_{\mathbf z\mathbf z}\mathcal L(\mathbf z^\ast)\mathbf e_N$ remains; components of $\mathbf e_N$ normal to an active inequality generate the first-order drift. In particular, the burnout dynamic-pressure drift $q_{b1}(\hat{\mathbf z})-q_{b1}(\mathbf z^\ast)\approx\nabla q_{b1}(\mathbf z^\ast)^T\mathbf e_N$ scales linearly with $\mathbf e_N$, providing the theoretical justification for incorporating the operational buffer $q_{tol}$.

In summary, propellant optimality is inherited directly from the learned initialization: the penalty scales quadratically with network error when inequality constraints remain inactive or locked, and exhibits at most linear drift proportional to the KKT multipliers when active inequalities are perturbed. Section~\ref{sec:accuracy} validates both theoretical regimes across the 3,000 test missions.

\section{Numerical Experiments}\label{sec:experiments}

\subsection{Setup: missions, methods and metrics}\label{sec:setup}

\paragraph{Test missions} All evaluations are conducted on the 3,000 held-out test missions from the SQP database, none of which were accessed during network training or hyperparameter selection. Comparisons against SLSQP and non-learned initializations use the first 1,000 test missions in their fixed sampling order; on this 1,000-mission subset, the success rate and propellant penalty of the proposed method closely match those across the full 3,000-mission test set.

\paragraph{Compared methods} All evaluated methods share the exact same event-aligned trajectory integrator:
\begin{itemize}
\item \emph{ResNet+SVD}: The proposed framework (Section~\ref{sec:method}).
\item \emph{SLSQP (cold, KKT)} and \emph{SLSQP (ResNet, KKT)}: SciPy's SLSQP optimizer solving Problem~(P) to standard KKT optimality ($ftol=10^{-9}$, central finite-difference gradients, $\bar q=400$~Pa limit, and a 30-iteration budget), initialized either from the standard engineering guess with $A_0=\pi/2+i^\star$ or directly from the ResNet prediction $\mathbf z_0$. These serve as the propellant-optimality reference and full online optimization baselines.
\item \emph{SLSQP (cold, early exit)} and \emph{SLSQP (ResNet, early exit)}: Equal-tolerance online baselines that monitor every trajectory evaluation (including line-search trial points) inside SLSQP and terminate immediately upon satisfying the orbital tolerances in~\eqref{eq:accuracy}, box bounds, and $q_{b1}\le400.01$~Pa.
\item \emph{Newton--GMRES}: A Jacobian-free Newton--Krylov baseline \cite{saad1986gmres,knoll2004jfnk,wang2018fqlnk} solving $\mathbf r(\mathbf z)=\mathbf 0$ from $\mathbf z_0$. Each solver instance updates a four-parameter subset (freezing the remaining four at $\mathbf z_0$) using at most four Arnoldi vectors per Newton step, a backtracking trust-region filter, and a 15-iteration cap; directional derivatives $\mathbf J\mathbf q_j$ use forward differences (one simulation) and switch to central differences (two concurrent simulations) when the step is small. Because a single four-parameter subset has a relatively low success rate (for example, $(\phi_1,\psi_2,\dot\phi_2,T_2)$ alone succeeds on only 91.6\% of the 1,000 test missions), four instances run concurrently on $(\phi_1,\psi_2,\dot\phi_2,T_2)$, $(A_0,\phi_2,T_2,\alpha_m)$, $(A_0,\phi_1,\phi_2,T_2)$, and $(\psi_2,\dot\phi_2,t_c,\alpha_m)$, synchronizing after each Newton step and terminating as soon as any instance meets~\eqref{eq:accuracy} (dispatching four or eight simulations per round; Fig.~\ref{fig:architecture}(b)). On a single core, instances are timed individually and we report four times the runtime of the fastest converging instance.
\item \emph{NN+SVD} and \emph{Default+SVD}: Ablation baselines running Algorithm~\ref{alg:svd} from non-learned initializations---specifically, a nearest-neighbor (NN) lookup in the normalized mission space over the 24,000 training labels, and the fixed engineering default used by SLSQP (cold). For both baselines, the iteration cap is relaxed to 100 to prevent premature termination from a tight iteration budget.
\end{itemize}

\paragraph{Evaluation metrics} A run is deemed \emph{successful} if the returned parameter vector satisfies the orbital tolerances in~\eqref{eq:accuracy}, respects all box bounds, and satisfies $q_{b1}\le\bar q+q_{tol}=415$~Pa when re-simulated using the fine 0.00625~s integration step. Propellant suboptimality is quantified by the relative burn-time excess~\eqref{eq:fuel} with respect to the database reference solution $\mathbf z^\ast$ obtained by SLSQP. Because $\mathbf z^\ast$ satisfies the KKT conditions of Section~\ref{sec:penalty} up to numerical tolerances, individual refined solutions can occasionally exhibit a slight negative penalty. Computational cost is measured by total simulations $N_{\mathrm{sim}}$, sequential rounds $N_{\mathrm{rnd}}$, and wall-clock execution time.

\paragraph{Computing platforms} Offline development and server benchmarks used an Intel Xeon Gold 6330 processor (2.0~GHz). The complete online pipeline is implemented in a self-contained C++ executable using only the header-only Eigen library without file I/O, compiling unchanged on both the server and the flight processor (Section~\ref{sec:onboard}). Server timings dispatch round simulations across $P$ threads and measure end-to-end wall-clock time from network input to final output. On eight server cores, SLSQP times are directly measured via SciPy with a multi-threaded C++ simulation bridge (reported as the mean of two runs over the first 1,000 test missions). On the bare-metal flight processor (where legacy Fortran SLSQP cannot run) and for the 1-core server baseline, SLSQP times are conservative lower-bound estimates multiplying exact SciPy evaluation counts by measured single-core simulation time, omitting QP subproblem overhead.

\subsection{Accuracy, dynamic pressure and propellant}\label{sec:accuracy}

Across all 3,000 test missions, ResNet+SVD achieves a 100\% success rate with a maximum burnout dynamic pressure of 406.4~Pa (well below the 415~Pa ceiling), whereas unrefined ResNet predictions satisfy the tolerances in zero missions. Table~\ref{tab:accuracy} reports terminal orbital errors across the first 1,000 test missions: ResNet+SVD, Newton--GMRES, and ResNet-initialized SLSQP all achieve 100\% success, while cold-started SLSQP fails on one mission. Both early-exit SLSQP and ResNet+SVD reliably satisfy all operational insertion tolerances, whereas KKT-terminated SLSQP iterates further to millimeter-level residuals.

\begin{table*}[t]
\centering
\caption{Terminal orbital errors (median and 95th percentile) across the first 1,000 test missions (successful runs only), re-verified using the fine 0.00625~s step. Target tolerances are 30~m, $10^{-6}$, and $10^{-6}$~rad. Med.: median; P95: 95th percentile.}\label{tab:accuracy}
\small
\begin{tabular}{lrrrrrrr}
\hline
Method & Success & \multicolumn{2}{c}{$|\Delta a|$ (m)} & \multicolumn{2}{c}{$|\Delta e|$ ($10^{-6}$)} & \multicolumn{2}{c}{$|\Delta i|$ ($10^{-6}$ rad)}\\
 & (\%) & Med. & P95 & Med. & P95 & Med. & P95\\
\hline
SLSQP (cold, KKT) & 99.9 & 0.006 & 0.064 & 0.002 & 0.017 & $<0.001$ & $<0.001$\\
SLSQP (cold, early exit) & 99.9 & 0.47 & 3.25 & 0.12 & 0.88 & 0.001 & 0.020\\
SLSQP (ResNet, KKT) & 100 & 0.006 & 0.075 & 0.002 & 0.019 & $<0.001$ & $<0.001$\\
SLSQP (ResNet, early exit) & 100 & 0.94 & 3.34 & 0.25 & 0.89 & 0.003 & 0.098\\
Newton--GMRES & 100 & 0.013 & 3.2 & 0.004 & 0.84 & $<0.001$ & 0.68\\
ResNet+SVD & 100 & 0.012 & 2.6 & 0.004 & 0.77 & $<0.001$ & 0.67\\
\hline
\end{tabular}
\end{table*}

Across all 3,000 test missions, whose database reference solutions all strictly satisfy~\eqref{eq:accuracy}, the magnitude of the relative propellant penalty $|\Delta m_{p2}|/m_{p2}^\ast$ achieves a median of 0.0010\%, a 90th percentile of 0.0045\%, and a 95th percentile of 0.0065\%. For a second stage carrying 80~kg of propellant, a 0.0065\% deviation corresponds to approximately 5~g of propellant.

To validate the theoretical regimes of Section~\ref{sec:penalty}, Table~\ref{tab:penalty} partitions the propellant penalty according to whether the dynamic-pressure limit is active at the reference optimum $\mathbf z^\ast$.

\begin{table}[t]
\centering
\caption{Relative propellant penalty of ResNet+SVD with respect to the SQP reference, partitioned by dynamic-pressure activity at $\mathbf z^\ast$ (first 1,000 test missions).}\label{tab:penalty}
\footnotesize
\setlength{\tabcolsep}{2.5pt}
\begin{tabular}{lrrrr}
\hline
Reference regime & Missions & \multicolumn{3}{c}{$|\Delta m_{p2}|/m_{p2}^\ast$ (\%)}\\
 & & Median & 95th pct. & Max.\\
\hline
$q_{b1}<399.9$~Pa (inactive) & 657 & 0.00077 & 0.0020 & 0.0059\\
$q_{b1}\ge399.9$~Pa (active) & 343 & 0.0027 & 0.0095 & 0.054\\
All & 1000 & 0.0010 & 0.0064 & 0.054\\
\hline
\end{tabular}
\end{table}

\begin{figure*}[t]
\centering
\includegraphics[width=\textwidth]{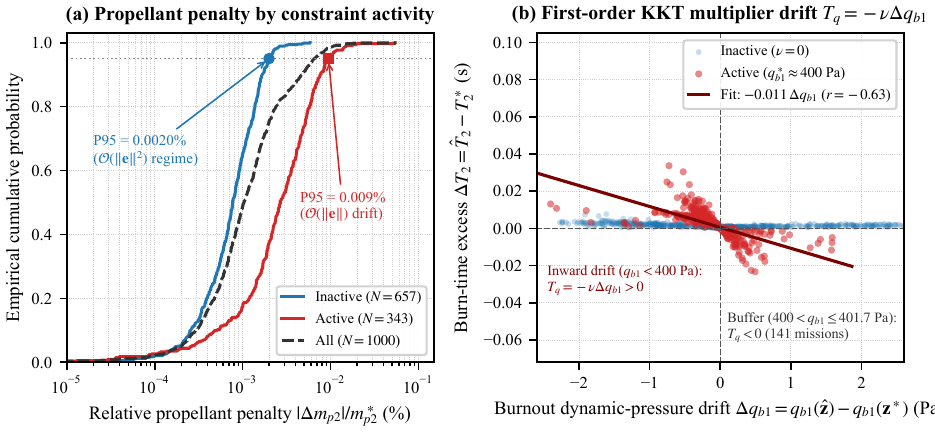}
\caption{Empirical verification of Theorem~\ref{thm:penalty} and Corollary~\ref{cor:order} across the first 1,000 test missions. (a)~Empirical cumulative distributions of the relative propellant penalty $|\Delta m_{p2}|/m_{p2}^\ast$ partitioned by whether the burnout dynamic-pressure limit is inactive ($q_{b1}(\mathbf z^\ast)<399.9$~Pa) or active ($q_{b1}(\mathbf z^\ast)\ge399.9$~Pa) at the SQP reference. (b)~Burn-time excess $\Delta T_2$ versus burnout dynamic-pressure drift $\Delta q_{b1}$, illustrating the first-order KKT multiplier slope $T_q=-\nu\Delta q_{b1}$ in the active regime alongside the negligible second-order residual in the inactive regime.}\label{fig:penalty}
\end{figure*}

Across all 3,000 test missions, $|\Delta m_{p2}|/m_{p2}^\ast$ achieves a median of 0.0010\%, a 90th percentile of 0.0045\%, and a 95th percentile of 0.0065\% ($\approx 5$~g for an 80~kg stage). Partitioning by dynamic-pressure activity at $\mathbf z^\ast$ (Table~\ref{tab:penalty}, Fig.~\ref{fig:penalty}) confirms Corollary~\ref{cor:order}: inactive-regime missions exhibit quadratic $O(\|\mathbf e\|^2)$ scaling ($\Delta T_2\approx 0$), whereas boundary-active missions show a fivefold larger 95th-percentile penalty and a negative linear correlation between $\Delta T_2$ and $\Delta q_{b1}$ (Pearson $-0.63$, regression slope $-0.011$~s/Pa matching $-\nu$ in Theorem~\ref{thm:penalty}). Dynamic-pressure drift stays within 1.0~Pa at the 95th percentile (at most $+6.4$~Pa across all 3,000 missions), and trajectories slightly entering the $q_{tol}$ buffer achieve marginally shorter burn times than the 400~Pa reference. ResNet-initialized Newton--GMRES and early-exit SLSQP yield virtually identical propellant statistics (0.0012\% and 0.0032\% medians; 0.0075\% and 0.0062\% 95th percentiles), confirming Theorem~\ref{thm:penalty} that near-optimality is governed by proximity to $\mathbf z^\ast$ rather than the specific solver.

\subsection{Simulation count, sequential rounds and wall-clock time}\label{sec:speed}

Table~\ref{tab:cost} compares computational complexity and wall-clock runtimes across the first 1,000 test missions.

\begin{table*}[t]
\centering
\caption{Computational cost across the first 1,000 test missions (mean, with 95th percentile in parentheses where available). Wall-clock times are measured on 1 and 8 cores of the Xeon Gold 6330 server (single-simulation runtime: 23.4~ms). For SLSQP, 8-core times are directly measured (mean over two runs on the first 1,000 test missions), whereas 1-core times are conservative lower-bound estimates omitting QP overhead. All methods achieve 100\% success except the cold-started SLSQP variants (99.9\%). $^{a}$~Total simulations across all four concurrent Newton--GMRES instances on 8 cores; on 1 core, four times the count of the fastest converging instance, 38.8 (48).}\label{tab:cost}
\small
\begin{tabular}{lrrrr}
\hline
Method & $N_{\mathrm{sim}}$ & $N_{\mathrm{rnd}}$ & 1 core (ms) & 8 cores (ms)\\
\hline
SLSQP (cold, KKT) & 396.4 (622) & 66.9 (172) & 9253 (14522) & 2570\\
SLSQP (cold, early exit) & 271.0 (307) & 32.8 (37) & 6341 (7184) & 1421\\
SLSQP (ResNet, KKT) & 399.8 (688) & 86.6 (238) & 9334 (16065) & 3137\\
SLSQP (ResNet, early exit) & 46.2 (69) & 6.3 (9) & 1081 (1615) & 267\\
Newton--GMRES & 41.6$^{a}$ (64) & 11.9 (28) & 909 (1183) & 278\\
ResNet+SVD & 31.9 (40) & 3.3 (4) & 746 (984) & 121\\
\hline
\end{tabular}
\end{table*}

Relative to SLSQP (cold, KKT), ResNet+SVD cuts simulations by $12\times$ and sequential rounds by $20\times$, running $12.4\times$ faster on one core and $21.3\times$ faster on eight cores (with $6.2\times$ vs.\ $3.6\times$ parallel scaling, as standard SLSQP averages 22.0 outer SQP iterations and 44.9 serial line-search trials). Relative to Newton--GMRES---which averages 9.7 simulations for its fastest instance (38.8 on one core vs.\ 41.6 across all four concurrent instances) but requires 11.9 sequential rounds (28 at the 95th percentile) due to its four-round Arnoldi orthogonalization chain per Newton step---ResNet+SVD is $1.2\times$ faster on one core and $2.3\times$ faster on eight cores, trading extra parallelizable evaluations to compress serial depth.

Comparing the KKT and early-exit SLSQP variants isolates the effects of formulation and stopping criteria. Under full KKT convergence, ResNet initialization does not accelerate SLSQP because it still constructs its BFGS Hessian from scratch near $\mathbf z^\ast$. Under early exit upon entering~\eqref{eq:accuracy}, ResNet initialization cuts SLSQP's simulations by 83\% and sequential rounds by 81\% relative to cold early exit. Nevertheless, ResNet+SVD still uses 31\% fewer simulations and nearly halves the sequential rounds ($3.3$ vs.\ $6.3$; $4$ vs.\ $9$ at the 95th percentile), running $1.4\times$ faster on one core and $2.2\times$ faster on eight server cores ($11.8\times$ faster than cold early-exit SLSQP). This remaining gap stems from problem formulation: warm-started SLSQP still solves a coupled constrained optimization problem from an identity BFGS Hessian, averaging 2.66 outer iterations (vs.\ 1.65 for minimum-norm feasibility restoration), so even parallelizing its 3.67 line-search evaluations would still leave five to six sequential rounds. Because both methods operate under the exact same operational tolerance box~\eqref{eq:accuracy} and satisfy all insertion accuracy requirements, this comparison is conducted under strictly identical stopping tolerances.

\subsection{Role of the learned initial guess}\label{sec:coldstart}

\begin{figure*}[t]
\centering
\includegraphics[width=\textwidth]{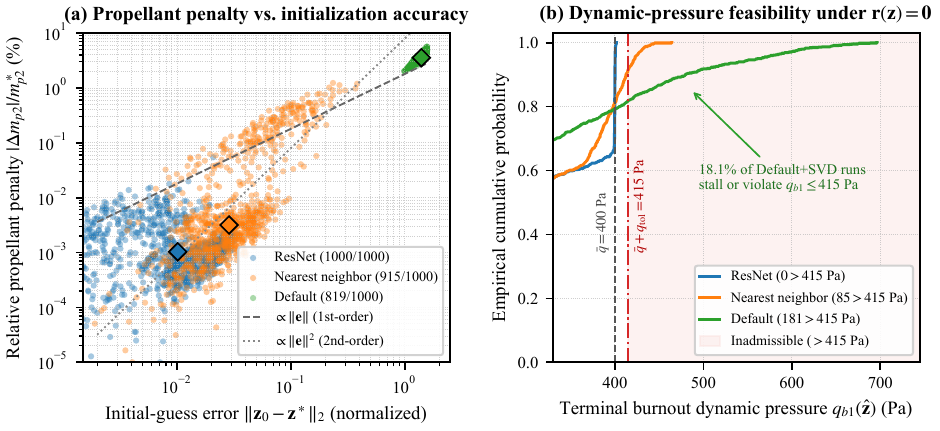}
\caption{Impact of initialization accuracy across the first 1,000 test missions. (a)~Relative propellant penalty $|\Delta m_{p2}|/m_{p2}^\ast$ versus normalized initial-guess error $\|\mathbf z_0-\mathbf z^\ast\|_2$ on successful runs (diamonds indicate group medians), shown alongside first- and second-order scaling references. (b)~Empirical cumulative distributions of terminal burnout dynamic pressure $q_{b1}(\hat{\mathbf z})$ when solving the terminal equations $\mathbf r(\mathbf z)=\mathbf 0$ without the admissibility filter.}\label{fig:coldstart}
\end{figure*}

Table~\ref{tab:coldstart} and Fig.~\ref{fig:coldstart} evaluate the refinement when initialized without the learned predictor.

\begin{table}[t]
\centering
\caption{Performance of Algorithm~\ref{alg:svd} under different initializations across the first 1,000 test missions. Distance denotes the normalized Euclidean error $\|\mathbf z_0-\mathbf z^\ast\|$; simulation counts and propellant penalties are reported over successful runs.}\label{tab:coldstart}
\footnotesize
\setlength{\tabcolsep}{2.5pt}
\begin{tabular}{lrrrrrr}
\hline
Initialization & Dist. & Succ. & $N_{\mathrm{sim}}$ & \multicolumn{3}{c}{$|\Delta m_{p2}|/m_{p2}^\ast$ (\%)}\\
 & (med.) & (/1000) & & Med. & P95 & Max.\\
\hline
ResNet & 0.0102 & 1000 & 31.9 & 0.0010 & 0.0064 & 0.054\\
Nearest neighbor & 0.031 & 915 & 38.7 & 0.0032 & 0.39 & 1.23\\
Eng.\ default & 1.43 & 819 & 183.7 & 3.56 & 4.78 & 5.68\\
\hline
\end{tabular}
\end{table}

As shown in Fig.~\ref{fig:coldstart}(a), across three orders of magnitude in $\|\mathbf e\|$, the propellant penalty follows the $\propto\|\mathbf e\|^2$ (inactive) and $\propto\|\mathbf e\|$ (active) envelopes of Corollary~\ref{cor:order}. Without a near-optimal initialization, Algorithm~\ref{alg:svd} degrades sharply (Table~\ref{tab:coldstart}): from the engineering default ($\|\mathbf e\|\approx 1.43$), 181 of 1,000 missions stall against the dynamic-pressure limit (reaching 696~Pa when the filter is removed; Fig.~\ref{fig:coldstart}(b)), while feasible runs incur a 3.56\% median penalty. Nearest-neighbor lookup avoids large median penalties but stalls in 8.5\% of missions, suffers a $60\times$ larger 95th-percentile penalty, and requires $3.36\times10^5$ stored entries versus $1.35\times10^5$ ResNet parameters. Thus, while SVD refinement eliminates terminal errors, propellant near-optimality and dynamic-pressure feasibility are inherited from the learned initial guess.

\subsection{Serial timing on an aerospace processor}\label{sec:onboard}

The C++ program of Section~\ref{sec:setup} was executed bare-metal (without an operating system) on a 1.25~GHz TI TMS320C6678 digital signal processor across the first 20 test missions, reproducing the server simulation counts and 100\% success rate. Implementing multi-core dispatch on bare-metal DSP hardware requires substantial low-level embedded engineering---including inter-core synchronization, shared-memory partitioning, and cache-coherency management (or porting a real-time operating system)---which lies outside the scope of this algorithmic study. Accordingly, we measure single-core serial runtimes on the DSP to establish exact onboard compute cost without unverified parallel-efficiency assumptions, leaving multi-core parallel scaling to the server benchmarks in Section~\ref{sec:speed}. With one trajectory simulation taking 3.421~s, ResNet+SVD averages $104.0$~s per mission ($30.4$ simulations), running $13.4\times$ faster than SLSQP (cold, KKT), $8.9\times$ faster than SLSQP (cold, early exit), $1.5\times$ faster than SLSQP (ResNet, early exit), and $1.3\times$ faster than Newton--GMRES (Table~\ref{tab:onboard}).

\begin{table}[t]
\centering
\caption{Serial computation time on the aerospace processor (s per mission, mean over the first 20 test missions). Times for ResNet+SVD and Newton--GMRES are measured on one core; SLSQP times are conservative lower-bound estimates computed as SciPy evaluation counts multiplied by the measured single-simulation time (3.421~s).}\label{tab:onboard}
\small
\begin{tabular}{lr}
\hline
Method & 1 core (s)\\
\hline
SLSQP (cold, KKT) & 1393\\
SLSQP (cold, early exit) & 922\\
SLSQP (ResNet, KKT) & 1326\\
SLSQP (ResNet, early exit) & 160.4\\
Newton--GMRES & 131.3\\
ResNet+SVD & 104.0\\
\hline
Speedup vs.\ SLSQP (cold, KKT)$^{a}$ & 13.4\\
Speedup vs.\ SLSQP (ResNet, early exit)$^{b}$ & 1.5\\
\hline
\multicolumn{2}{p{0.85\columnwidth}}{\footnotesize $^{a,b}$ Mean time of the respective SLSQP baseline divided by ResNet+SVD.}
\end{tabular}
\end{table}

\section{Conclusion}\label{sec:conclusion}

This paper presented an optimization-free onboard ascent parameter determination framework pairing an offline-trained ResNet predictor with a parallel boundary-frozen damped SVD solver, achieving fast, high-precision, and near-propellant-optimal trajectory generation. Theoretical analysis establishes an explicit initial-error threshold for local convergence, shows that the minimum-norm SVD step restores terminal feasibility without injecting extraneous null-space drift, and proves via an exact KKT decomposition that propellant suboptimality scales quadratically with prediction error when inequalities are inactive or locked (and at most linearly otherwise). Across 3,000 test missions, the framework achieved 100\% constraint compliance with propellant within 0.0065\% of the KKT-optimal SQP reference in 95\% of cases, while achieving an $11.8\times$ eight-core server speedup over cold early-exit SLSQP ($2.2\times$ over warm early-exit SLSQP; $2.3\times$ over Newton--GMRES) and an $8.9\times$ serial speedup on a flight processor ($1.5\times$ and $1.3\times$, respectively). Future work will port the parallel dispatch layer to the multi-core flight processor for hardware-in-the-loop qualification, evaluate robustness under extreme dispersions, and extend the formulation to elliptical target orbits and active dynamic-pressure control.

\appendix

\section{Proofs for Section~\ref{sec:theory}}\label{app:proofs}

We adopt the notation established in Section~\ref{sec:theory}. At iteration $k$, let $\mathbf r_k=\mathbf r(\mathbf z_k)$, and let $\mathbf d_k$ denote the unscaled damped SVD step prior to trust-region clipping, padded with zeros on the frozen coordinates. (At the first iteration, replacing $\mathbf r_0=\mathbf r(\mathbf z_0)$ by the central-difference pair average $\bar{\mathbf r}$ introduces an $O(h^2)$ residual discrepancy, which simply augments $\chi$ at $k=0$ by the relative term $\|\bar{\mathbf r}-\mathbf r_0\|/\|\mathbf r_0\|$.) Because $\mathbf d_k$ perturbs only the free coordinates, we identify $\mathbf d_k$ with its free-coordinate subvector in $\mathbb R^m$, so that all Taylor expansions around $\mathbf z_k$ below involve exclusively the reduced Jacobian $\mathbf J_{\mathcal I}$.

\subsection{Proof of Lemma~\ref{lem:contraction}}\label{app:lemma1}

Let $\mathbf A=\tilde{\mathbf J}_k$ and $\mu=\mu_k$. By Assumptions~(A2) and~(A3) together with Weyl's inequality for singular values \cite{horn2012matrix}, every singular value of $\mathbf A$ satisfies $\sigma_i\ge\sigma-\varepsilon_J=\tilde\sigma>0$. The damped SVD step is given by
\begin{equation*}
\begin{aligned}
\mathbf d_k&=-\mathbf A^T\left(\mathbf A\mathbf A^T+\mu^2\mathbf I\right)^{-1}\mathbf r_k\\
&=-\mathbf V\,\mathrm{diag}\!\left(\frac{\sigma_i}{\sigma_i^2+\mu^2}\right)\mathbf U^T\mathbf r_k .
\end{aligned}
\end{equation*}
Because $\sigma_i/(\sigma_i^2+\mu^2)\le1/\sigma_i\le1/\tilde\sigma$, the step-norm bound $\|\mathbf d_k\|\le\|\mathbf r_k\|/\tilde\sigma$ follows immediately.

For the residual contraction bound, convexity of $\mathcal B$ ensures that the line segment connecting $\mathbf z_k$ and $\mathbf z_k+\mathbf d_k$ lies entirely within $\mathcal B$. By Assumption~(A1), expanding $\mathbf r$ around $\mathbf z_k$ yields
\begin{equation*}
\mathbf r(\mathbf z_k+\mathbf d_k)=\mathbf r_k+\mathbf J_{\mathcal I}(\mathbf z_k)\mathbf d_k+\mathbf R_{r,k},\quad \|\mathbf R_{r,k}\|\le\tfrac L2\|\mathbf d_k\|^2 .
\end{equation*}
We decompose the linear terms as $\mathbf r_k+\mathbf J_{\mathcal I}(\mathbf z_k)\mathbf d_k=(\mathbf r_k+\mathbf A\mathbf d_k)+(\mathbf J_{\mathcal I}(\mathbf z_k)-\mathbf A)\mathbf d_k$. Substituting $\mathbf d_k$ into the first term yields
\begin{equation*}
\mathbf r_k+\mathbf A\mathbf d_k=\mu^2\left(\mathbf A\mathbf A^T+\mu^2\mathbf I\right)^{-1}\mathbf r_k ,
\end{equation*}
whose norm is bounded by $\frac{\mu^2}{\tilde\sigma^2+\mu^2}\|\mathbf r_k\|\le(\bar\mu/\tilde\sigma)^2\|\mathbf r_k\|$. By Assumption~(A3), the second term is bounded in norm by $\varepsilon_J\|\mathbf d_k\|\le(\varepsilon_J/\tilde\sigma)\|\mathbf r_k\|$. Finally, the quadratic remainder satisfies $\|\mathbf R_{r,k}\|\le\tfrac L2\|\mathbf r_k\|^2/\tilde\sigma^2$. Combining these three bounds via the triangle inequality gives
\begin{equation*}
\|\mathbf r(\mathbf z_k+\mathbf d_k)\|\le\left(\chi+\frac{L\|\mathbf r_k\|}{2\tilde\sigma^2}\right)\|\mathbf r_k\| .
\end{equation*}
\hfill$\square$

\subsection{Proof of Theorem~\ref{thm:convergence}}\label{app:thm1}

We prove by induction on $k\ge0$ that
\begin{equation}
\mathbf z_k\in\mathcal B,\quad \|\mathbf r_k\|\le2^{-k}R_0,\quad \|\mathbf z_k-\mathbf z_0\|\le\sum_{j=0}^{k-1}2^{-j}\frac{R_0}{\tilde\sigma}.
\label{eq:induction}
\end{equation}
The base case $k=0$ holds trivially (with the empty sum equal to zero). Suppose~\eqref{eq:induction} holds at iteration $k$.

\emph{Inclusion of the full step in $\mathcal B$.} By the step bound in Lemma~\ref{lem:contraction} and the induction hypothesis, $\|\mathbf d_k\|\le\|\mathbf r_k\|/\tilde\sigma\le2^{-k}R_0/\tilde\sigma$. Consequently, $\|\mathbf z_k+\mathbf d_k-\mathbf z_0\|\le\sum_{j=0}^{k}2^{-j}R_0/\tilde\sigma<2R_0/\tilde\sigma\le r_B$ by~\eqref{eq:thm1cond}, ensuring $\mathbf z_k+\mathbf d_k\in\mathcal B$ so that Lemma~\ref{lem:contraction} applies.

\emph{Inactivity of trust-region scaling.} Because the trust radius resets to $\rho_0$ after every accepted step (and we establish below that every step is accepted), condition~\eqref{eq:thm1cond} ensures $\|\mathbf d_k\|\le R_0/\tilde\sigma\le\rho_0$. Thus, trust-region clipping remains inactive.

\emph{Residual halving by the full step.} Since $\|\mathbf r_k\|\le R_0\le\tilde\sigma^2/(2L)$, we have $L\|\mathbf r_k\|/(2\tilde\sigma^2)\le1/4$. Combining this with $\chi\le1/4$ from Assumption~(A4) in Lemma~\ref{lem:contraction} yields $\|\mathbf r(\mathbf z_k+\mathbf d_k)\|\le\tfrac12\|\mathbf r_k\|$.

\emph{Step acceptance.} By Assumption~(A5), the projection $\Pi(\cdot)$ leaves all candidate iterates in $\mathcal B$ unaltered, no additional coordinates are frozen, and all candidates in $\mathcal B$ are admissible. If any candidate satisfies the terminal orbital tolerances~\eqref{eq:accuracy}, the algorithm terminates successfully at an admissible solution. Otherwise, because the full step ($\alpha=1$) is admissible and strictly reduces the residual norm, Assumption~(A5) guarantees that the accepted iterate $\mathbf z_{k+1}=\mathbf z_k+\alpha_k\mathbf d_k$ satisfies $\|\mathbf r_{k+1}\|\le\|\mathbf r(\mathbf z_k+\mathbf d_k)\|\le\tfrac12\|\mathbf r_k\|\le2^{-(k+1)}R_0$. Since $\alpha_k\le1$, we also have $\|\mathbf z_{k+1}-\mathbf z_k\|\le\|\mathbf d_k\|\le2^{-k}R_0/\tilde\sigma$, completing the induction.

\emph{Asymptotic convergence.} From~\eqref{eq:induction}, $\|\mathbf z_k-\mathbf z_0\|\le2R_0/\tilde\sigma$ holds for all $k$, and $\|\mathbf r_k\|\le\tau$ is achieved within at most $\lceil\log_2(R_0/\tau)\rceil$ iterations. Because the step displacements $\|\mathbf z_{k+1}-\mathbf z_k\|$ are summable, $\{\mathbf z_k\}$ forms a Cauchy sequence converging to a limit $\mathbf z_\infty$. Continuity of $\mathbf r$ implies $\mathbf r(\mathbf z_\infty)=\mathbf 0$, while closedness of $\mathcal B$ and Assumption~(A5) ensure $\mathbf z_\infty\in\mathcal B$ and strict satisfaction of all box bounds.

\emph{Two-stage contraction rate.} Let $\zeta_k=L\|\mathbf r_k\|/(2\tilde\sigma^2)$ denote the normalized residual. By Lemma~\ref{lem:contraction}, $\zeta_{k+1}\le(\chi+\zeta_k)\zeta_k$. Thus, for every iteration satisfying $\zeta_k\ge\chi$, the residual contracts quadratically according to $\zeta_{k+1}\le2\zeta_k^2$, or equivalently,
\begin{equation*}
\|\mathbf r_{k+1}\|\le\frac{L}{\tilde\sigma^2}\|\mathbf r_k\|^2;\quad\text{once }\zeta_k<\chi,\ \ \|\mathbf r_{k+1}\|\le2\chi\|\mathbf r_k\|.
\end{equation*}
\hfill$\square$

\subsection{Proof of Corollary~\ref{cor:emax}}\label{app:cor1}

Because the line segment between $\mathbf z^\ast$ and $\mathbf z_0=\mathbf z^\ast+\mathbf e$ lies in $\mathcal U$ and $\mathbf r(\mathbf z^\ast)=\mathbf 0$, the mean-value inequality yields
\begin{equation*}
R_0=\|\mathbf r(\mathbf z^\ast+\mathbf e)-\mathbf r(\mathbf z^\ast)\|\le\sup_{\mathbf z\in\mathcal U}\|\mathbf J(\mathbf z)\|\,\|\mathbf e\|=M_J\|\mathbf e\| .
\end{equation*}
Because this bound employs the full eight-column Jacobian $\mathbf J$, it accounts for initial prediction errors across both free and frozen coordinates. Whenever $\|\mathbf e\|\le e_{\max}$ as defined in~\eqref{eq:emax}, the bound $R_0\le M_J e_{\max}$ fulfills all three conditions in~\eqref{eq:thm1cond}, while $\|\mathbf z_0-\mathbf z^\ast\|\le\bar e$ ensures that the uniform constants in Assumptions~(A1)--(A5) apply. Because the mapping from the terminal state to $(a-a^\star,e,i-i^\star)$ is locally Lipschitz continuous near the target circular orbit and vanishes if and only if $\mathbf r(\mathbf z)=\mathbf 0$, the orbital accuracy requirements~\eqref{eq:accuracy} are implied by $\|\mathbf r(\mathbf z)\|\le\tau_{\mathrm{acc}}$ for a suitable threshold $\tau_{\mathrm{acc}}>0$. Theorem~\ref{thm:convergence} then guarantees convergence within at most $\lceil\log_2(R_0/\tau_{\mathrm{acc}})\rceil$ iterations (assuming an adequate iteration budget) to a point $\hat{\mathbf z}$ that satisfies~\eqref{eq:accuracy}, respects all box bounds, and---by Assumption~(A5)---satisfies the dynamic-pressure limit, thereby solving Problem~(F). Finally, the triangle inequality gives
\begin{equation*}
\begin{aligned}
\|\hat{\mathbf z}-\mathbf z_0\|&\le\frac{2R_0}{\tilde\sigma}\le\frac{2M_J}{\tilde\sigma}\|\mathbf e\|,\\
D=\|\hat{\mathbf z}-\mathbf z^\ast\|&\le\|\mathbf e\|+\|\hat{\mathbf z}-\mathbf z_0\|\le c_e\|\mathbf e\| .
\end{aligned}
\end{equation*}
\hfill$\square$

\subsection{Proof of Corollary~\ref{cor:order}}\label{app:cor2}

From the exact decomposition~\eqref{eq:penalty} of Theorem~\ref{thm:penalty}, subtracting the first-order inequality drift terms yields
\begin{equation*}
\begin{aligned}
\left|\Delta T_2-T_q-T_B\right|&=\left|-\boldsymbol\lambda^T\mathbf r(\hat{\mathbf z})+\boldsymbol\lambda^T\mathbf R_r+\nu R_g\right|\\
&\le\|\boldsymbol\lambda\|\,\|\mathbf r(\hat{\mathbf z})\|+\|\boldsymbol\lambda\|\,\|\mathbf R_r\|+\nu|R_g| .
\end{aligned}
\end{equation*}
Substituting $\|\mathbf r(\hat{\mathbf z})\|\le\varepsilon_c$, the Taylor bounds $\|\mathbf R_r\|\le\tfrac12 LD^2$ and $|R_g|\le\tfrac12 L_gD^2$, and the displacement bound $D\le c_e\|\mathbf e\|$ from Corollary~\ref{cor:emax} gives the first inequality directly.

For the dynamic-pressure term $T_q=-\nu g(\hat{\mathbf z})$, admissibility in Assumption~(A5) requires $q_{b1}(\hat{\mathbf z})\le\bar q+q_{tol}$, whence $g(\hat{\mathbf z})\le q_{tol}$ and $T_q\ge-\nu q_{tol}$. For the upper bound, Taylor expansion of $g$ around $\mathbf z^\ast$ with $\nu g(\mathbf z^\ast)=0$ gives
\begin{equation*}
\begin{aligned}
-g(\hat{\mathbf z})&=-\nabla g(\mathbf z^\ast)^T\boldsymbol\delta-R_g\le G_q D+\tfrac12 L_g D^2\\
&\le G_q c_e\|\mathbf e\|+\tfrac12 L_g c_e^2\|\mathbf e\|^2 ,
\end{aligned}
\end{equation*}
which yields the upper bound upon multiplication by $\nu\ge0$. Finally, for the parameter bounds, box-constraint satisfaction $\hat z_j\in[-1,1]$ implies $b_j(\hat{\mathbf z})\ge0$ and hence $T_B\ge0$. Because each active bound satisfies $b_j(\hat{\mathbf z})=|z^\ast_j-\hat z_j|=|\delta_j|\le D\le c_e\|\mathbf e\|$, the upper bound on $T_B$ follows immediately.
\hfill$\square$

\subsection{Proof of Proposition~\ref{prop:tangent}}\label{app:prop1}

Throughout this proof, let $\varepsilon=\|\mathbf e\|\le e_{\max}$; all generic constants $C$ depend solely on $\tilde\sigma$, $M_J$, $L$, $L_g$, $G_q$, $L_2$, $L_{g,2}$, $\bar e$, and the KKT multipliers $(\boldsymbol\lambda,\nu,\eta_j)$, and are independent of $\mathbf e$. Here, $L_2$ and $L_{g,2}$ denote the Lipschitz constants of the Hessians of $\mathbf r$ and $g$ on $\mathcal U$. Letting $\mathbf H_r(\mathbf z)[\mathbf u,\mathbf u]\in\mathbb R^4$ denote the second directional derivative of $\mathbf r$ along $\mathbf u$, we define
\begin{equation*}
\begin{aligned}
\mathbf H&=\nabla^2_{\mathbf z\mathbf z}\mathcal L(\mathbf z^\ast)=\sum_{i=1}^4\lambda_i\nabla^2r_i(\mathbf z^\ast)+\nu\nabla^2g(\mathbf z^\ast),\\
\bar H&=\|\boldsymbol\lambda\|L+\nu L_g .
\end{aligned}
\end{equation*}
Because $\mathbf J$ is $L$-Lipschitz continuous, we have $\|\mathbf H_r(\mathbf z)[\mathbf u,\mathbf u]\|\le L\|\mathbf u\|^2$; similarly, the $L_g$-Lipschitz continuity of $\nabla g$ implies $\|\nabla^2g\|\le L_g$. Since $\nabla^2f=\mathbf 0$, the symmetric Hessian $\mathbf H$ satisfies $\|\mathbf H\|=\sup_{\|\mathbf u\|=1}|\mathbf u^T\mathbf H\mathbf u|\le\bar H$.

\emph{Preliminary observations.} (i)~In the absence of frozen coordinates, $\mathcal I=\{1,\dots,8\}$, $\mathbf J_{\mathcal I}=\mathbf J$, and $\mathcal B$ reduces to the ball of radius $r_B$ centered at $\mathbf z_0$ intersected with $\mathcal U$. As established in the proof of Corollary~\ref{cor:emax}, $\mathbf z_0\in\mathcal B$, the iterates of Algorithm~\ref{alg:svd} satisfy the induction hypotheses of Theorem~\ref{thm:convergence}, trust-region clipping remains inactive, and, by assumption, the first iteration accepts the unclipped full step $\mathbf z_1=\mathbf z_0+\mathbf d_0$. Moreover, the refined output $\hat{\mathbf z}$ lies in $\mathcal B\subset\mathcal U$, and by the convexity of $\mathcal U$, the line segment connecting $\mathbf z^\ast\in\mathcal U$ and $\hat{\mathbf z}$ is contained entirely within $\mathcal U$.
(ii)~Because an active box bound at $\mathbf z^\ast$ precludes the interior margin condition of Assumption~(A5), $\mathbf z^\ast$ need not lie in $\mathcal B$, so Assumption~(A2) cannot be invoked directly at $\mathbf z^\ast$. Instead, combining $\tilde\sigma\le\sigma\le\sigma_{\min}(\mathbf J(\mathbf z_0))\le M_J$ with $L\varepsilon\le Le_{\max}\le\tilde\sigma^2/(2M_J)\le\tilde\sigma/2$ yields, via Weyl's inequality,
\begin{equation}
\sigma_{\min}(\mathbf J^\ast)\ge\sigma_{\min}(\mathbf J(\mathbf z_0))-\|\mathbf J(\mathbf z_0)-\mathbf J^\ast\|\ge\sigma-L\varepsilon\ge\tilde\sigma/2>0 .
\label{eq:sigmastar}
\end{equation}

\paragraph{Step 1: Annihilation of the row-space error at the first step.}
Let $\mathbf A=\tilde{\mathbf J}_0$, $\mu=\mu_0$, and $\mathbf E=\mathbf A-\mathbf J^\ast=(\tilde{\mathbf J}_0-\mathbf J(\mathbf z_0))+(\mathbf J(\mathbf z_0)-\mathbf J^\ast)$, which satisfies $\|\mathbf E\|\le\varepsilon_J+L\varepsilon$. As in the proof of Lemma~\ref{lem:contraction}, every singular value of $\mathbf A$ is bounded below by $\tilde\sigma$. The residual employed at the first iteration is $\tilde{\mathbf r}_0=\mathbf r_0+\boldsymbol\xi$, where $\boldsymbol\xi=\bar{\mathbf r}-\mathbf r_0$; by the convention adopted at the beginning of this appendix, $\chi$ absorbs the relative discrepancy so that $\|\boldsymbol\xi\|\le\chi\|\mathbf r_0\|$. Expanding $\mathbf r$ around $\mathbf z^\ast$ via Taylor's theorem gives $\mathbf r_0=\mathbf J^\ast\mathbf e+\boldsymbol\rho_2$, with $\|\boldsymbol\rho_2\|\le\tfrac12L\varepsilon^2$ and $\|\mathbf r_0\|\le M_J\varepsilon$. Writing the SVD of $\mathbf A$ as $\mathbf A=\mathbf U\boldsymbol\Sigma\mathbf V^T$, we define
\begin{equation*}
\begin{aligned}
\mathbf B_\mu&=\mathbf A^T(\mathbf A\mathbf A^T+\mu^2\mathbf I)^{-1}=\mathbf V\,\mathrm{diag}\!\Big(\frac{\sigma_i}{\sigma_i^2+\mu^2}\Big)\mathbf U^T,\\
\mathbf d_0&=-\mathbf B_\mu\tilde{\mathbf r}_0.
\end{aligned}
\end{equation*}
Then $\|\mathbf B_\mu\|\le1/\tilde\sigma$, and
\begin{equation*}
\begin{aligned}
\mathbf B_\mu\mathbf A&=\mathbf V\,\mathrm{diag}\!\Big(\frac{\sigma_i^2}{\sigma_i^2+\mu^2}\Big)\mathbf V^T=\mathbf P_A-\mathbf D_\mu,\\
\mathbf D_\mu&=\mathbf V\,\mathrm{diag}\!\Big(\frac{\mu^2}{\sigma_i^2+\mu^2}\Big)\mathbf V^T,\qquad\|\mathbf D_\mu\|\le(\bar\mu/\tilde\sigma)^2,
\end{aligned}
\end{equation*}
where $\mathbf P_A=\mathbf V\mathbf V^T=\mathbf A^T(\mathbf A\mathbf A^T)^{-1}\mathbf A$ denotes the orthogonal projector onto $\mathrm{row}(\mathbf A)$. Substituting $\mathbf J^\ast=\mathbf A-\mathbf E$ yields
\begin{equation}
\begin{aligned}
-\mathbf d_0&=\mathbf B_\mu\mathbf J^\ast\mathbf e+\mathbf B_\mu(\boldsymbol\rho_2+\boldsymbol\xi)\\
&=\mathbf P_A\mathbf e-\mathbf D_\mu\mathbf e-\mathbf B_\mu\mathbf E\mathbf e+\mathbf B_\mu(\boldsymbol\rho_2+\boldsymbol\xi).
\end{aligned}
\label{eq:d0exp}
\end{equation}
To relate $\mathbf P_A\mathbf e$ to $\mathbf e_R=\mathbf P\mathbf e$, we decompose the projector difference as $\mathbf P_A-\mathbf P=\mathbf P_A(\mathbf I-\mathbf P)-(\mathbf I-\mathbf P_A)\mathbf P$. Because $(\mathbf I-\mathbf P)\mathbf J^{\ast T}=\mathbf 0$, we have $(\mathbf I-\mathbf P)\mathbf A^T=(\mathbf I-\mathbf P)\mathbf E^T$, and therefore
\begin{equation*}
\begin{aligned}
\|\mathbf P_A(\mathbf I-\mathbf P)\|&=\|(\mathbf I-\mathbf P)\mathbf P_A\|=\|(\mathbf I-\mathbf P)\mathbf E^T(\mathbf A\mathbf A^T)^{-1}\mathbf A\|\\
&\le\frac{\|\mathbf E\|}{\sigma_{\min}(\mathbf A)}\le\frac{\|\mathbf E\|}{\tilde\sigma}.
\end{aligned}
\end{equation*}
Analogously, $(\mathbf I-\mathbf P_A)\mathbf A^T=\mathbf 0$ implies $(\mathbf I-\mathbf P_A)\mathbf J^{\ast T}=-(\mathbf I-\mathbf P_A)\mathbf E^T$, which together with~\eqref{eq:sigmastar} gives
\begin{equation*}
\begin{aligned}
\|(\mathbf I-\mathbf P_A)\mathbf P\|&=\|(\mathbf I-\mathbf P_A)\mathbf E^T(\mathbf J^\ast\mathbf J^{\ast T})^{-1}\mathbf J^\ast\|\\
&\le\frac{\|\mathbf E\|}{\sigma_{\min}(\mathbf J^\ast)}\le\frac{2\|\mathbf E\|}{\tilde\sigma}.
\end{aligned}
\end{equation*}
Consequently, $\|\mathbf P_A\mathbf e-\mathbf e_R\|\le3\|\mathbf E\|\varepsilon/\tilde\sigma$. Combining this bound with the estimates for $\mathbf D_\mu$ and $\mathbf B_\mu$ in~\eqref{eq:d0exp}, and using $\|\boldsymbol\rho_2\|\le\tfrac12L\varepsilon^2$ and $\|\boldsymbol\xi\|\le\chi M_J\varepsilon$, we obtain
\begin{equation}
\begin{aligned}
\|\mathbf d_0+\mathbf e_R\|&\le\frac{4\|\mathbf E\|\varepsilon}{\tilde\sigma}+\Big(\frac{\bar\mu}{\tilde\sigma}\Big)^2\varepsilon+\frac{\tfrac12L\varepsilon^2+\chi M_J\varepsilon}{\tilde\sigma}\\
&\le\frac{9L}{2\tilde\sigma}\varepsilon^2+\Big(4+\frac{M_J}{\tilde\sigma}\Big)\chi\,\varepsilon ,
\end{aligned}
\label{eq:d0bound}
\end{equation}
where the second inequality follows from $4(\varepsilon_J+L\varepsilon)\varepsilon/\tilde\sigma+(\bar\mu/\tilde\sigma)^2\varepsilon\le4\chi\varepsilon+4L\varepsilon^2/\tilde\sigma$. (An analogous estimate holds for the damped step at any iteration, though only $k=0$ is required here.)

\paragraph{Step 2: Second-order bound on subsequent iterations.}
Applying Lemma~\ref{lem:contraction} (under the above convention for $\boldsymbol\xi$) together with $\|\tilde{\mathbf r}_0\|\le(1+\chi)M_J\varepsilon\le\tfrac54M_J\varepsilon$ yields
\begin{equation*}
\begin{aligned}
\|\mathbf r_1\|&\le\Big(\chi+\frac{L\|\tilde{\mathbf r}_0\|}{2\tilde\sigma^2}\Big)\|\tilde{\mathbf r}_0\|\\
&\le\tfrac54M_J\chi\,\varepsilon+\frac{25LM_J^2}{32\tilde\sigma^2}\varepsilon^2 .
\end{aligned}
\end{equation*}
By the induction established in the proof of Theorem~\ref{thm:convergence}, all subsequent iterations satisfy $\|\mathbf r_{k+1}\|\le\tfrac12\|\mathbf r_k\|$ and $\|\mathbf z_{k+1}-\mathbf z_k\|\le\|\mathbf r_k\|/\tilde\sigma$, and the returned solution $\hat{\mathbf z}$ is either an iterate or a candidate step $\mathbf z_K+\alpha\mathbf d_K$ with $\alpha\le1$. Consequently,
\begin{equation*}
\begin{aligned}
\|\hat{\mathbf z}-\mathbf z_1\|&\le\sum_{k\ge1}\frac{\|\mathbf r_k\|}{\tilde\sigma}\le\frac{2\|\mathbf r_1\|}{\tilde\sigma}\\
&\le\frac{5M_J}{2\tilde\sigma}\chi\,\varepsilon+\frac{25LM_J^2}{16\tilde\sigma^3}\varepsilon^2 .
\end{aligned}
\end{equation*}

\paragraph{Step 3: Final deviation and dynamic-pressure drift.}
Because $\boldsymbol\delta=\mathbf e+\mathbf d_0+(\hat{\mathbf z}-\mathbf z_1)$ and $\mathbf e-\mathbf e_N=\mathbf e_R$, the residual deviation $\mathbf s=\boldsymbol\delta-\mathbf e_N=(\mathbf d_0+\mathbf e_R)+(\hat{\mathbf z}-\mathbf z_1)$ is bounded via~\eqref{eq:d0bound} and Step~2 by
\begin{equation}
\begin{aligned}
&\|\mathbf s\|\le\omega:=K_1\varepsilon^2+K_2\chi\,\varepsilon,\\
&K_1=\frac{9L}{2\tilde\sigma}+\frac{25LM_J^2}{16\tilde\sigma^3},\qquad K_2=4+\frac{7M_J}{2\tilde\sigma}.
\end{aligned}
\label{eq:sbound}
\end{equation}
This establishes the first claim, $\boldsymbol\delta=\mathbf e_N+O(\varepsilon^2)+O(\chi)\varepsilon$. Expanding $g$ around $\mathbf z^\ast$ then gives $g(\hat{\mathbf z})-g(\mathbf z^\ast)=\nabla g(\mathbf z^\ast)^T\boldsymbol\delta+R_g=\nabla g(\mathbf z^\ast)^T\mathbf e_N+\nabla g(\mathbf z^\ast)^T\mathbf s+R_g$, where $|\nabla g(\mathbf z^\ast)^T\mathbf s|\le G_q\omega$ and $|R_g|\le\tfrac12L_gD^2\le\tfrac12L_gc_e^2\varepsilon^2$, establishing the second claim.

\paragraph{Step 4: Second-order expansion of the propellant penalty.}
Starting from the exact decomposition~\eqref{eq:penalty}, we retain $T_c=-\boldsymbol\lambda^T\mathbf r(\hat{\mathbf z})$, $T_q=-\nu g(\hat{\mathbf z})$, and $T_B$ in their exact forms. By the Lipschitz continuity of the Hessians, Taylor's theorem with integral remainder yields
\begin{equation*}
\begin{aligned}
\mathbf R_r&=\tfrac12\mathbf H_r(\mathbf z^\ast)[\boldsymbol\delta,\boldsymbol\delta]+\mathbf R_r^{(3)},\qquad \|\mathbf R_r^{(3)}\|\le\tfrac16L_2D^3,\\
R_g&=\tfrac12\boldsymbol\delta^T\nabla^2g(\mathbf z^\ast)\boldsymbol\delta+R_g^{(3)},\qquad |R_g^{(3)}|\le\tfrac16L_{g,2}D^3 .
\end{aligned}
\end{equation*}
Consequently, setting $Q(\mathbf u)=\tfrac12\mathbf u^T\mathbf H\mathbf u$ and noting that $\nabla^2f=\mathbf 0$, we obtain
\begin{equation*}
\begin{aligned}
T^{(2)}&=\boldsymbol\lambda^T\mathbf R_r+\nu R_g\\
&=Q(\boldsymbol\delta)+\boldsymbol\lambda^T\mathbf R_r^{(3)}+\nu R_g^{(3)},\\
\big|\boldsymbol\lambda^T\mathbf R_r^{(3)}+\nu R_g^{(3)}\big|&\le\tfrac16(\|\boldsymbol\lambda\|L_2+\nu L_{g,2})c_e^3\varepsilon^3 .
\end{aligned}
\end{equation*}
Using $\|\mathbf e_N\|\le\varepsilon$, the error incurred by replacing $\boldsymbol\delta=\mathbf e_N+\mathbf s$ with $\mathbf e_N$ in $Q$ is bounded by
\begin{equation*}
\begin{aligned}
|Q(\boldsymbol\delta)-Q(\mathbf e_N)|&=\tfrac12|\mathbf s^T\mathbf H(\boldsymbol\delta+\mathbf e_N)|\\
&\le\tfrac12\bar H\,\|\mathbf s\|\,(2\varepsilon+\|\mathbf s\|)\\
&\le\tfrac12\bar H\,\omega(2\varepsilon+\omega).
\end{aligned}
\end{equation*}
From~\eqref{eq:sbound}, we have $\omega\,\varepsilon=K_1\varepsilon^3+K_2\chi\varepsilon^2$; furthermore, since $\chi\le\frac14$ and $\varepsilon\le\bar e$, it follows that $\omega^2\le2K_1^2\varepsilon^4+2K_2^2\chi^2\varepsilon^2\le2K_1^2\bar e\,\varepsilon^3+\tfrac12K_2^2\chi\varepsilon^2$. Combining these estimates bounds the total error of approximating $T^{(2)}$ by $\tfrac12\mathbf e_N^T\mathbf H\mathbf e_N$ within $C(\varepsilon^3+\chi\varepsilon^2)$, which yields~\eqref{eq:tangent}. Equivalently, because $T_q+\nu R_g=-\nu\nabla g(\mathbf z^\ast)^T\boldsymbol\delta$ (as the second-order remainders $-\nu R_g$ and $+\nu R_g$ cancel), one may write
\begin{equation*}
\begin{aligned}
\Delta T_2={}&-\boldsymbol\lambda^T\mathbf r(\hat{\mathbf z})-\nu\nabla g(\mathbf z^\ast)^T\boldsymbol\delta+T_B\\
&+\tfrac12\mathbf e_N^T\Big(\sum_{i=1}^4\lambda_i\nabla^2r_i(\mathbf z^\ast)\Big)\mathbf e_N+O(\varepsilon^3+\chi\varepsilon^2).
\end{aligned}
\end{equation*}
In this alternative form, however, the linearized term $-\nu\nabla g(\mathbf z^\ast)^T\boldsymbol\delta$ retains a second-order contribution $-\nu\nabla g(\mathbf z^\ast)^T\mathbf s$ of the same asymptotic order as the quadratic curvature term; retaining the exact drift $-\nu g(\hat{\mathbf z})$ as in~\eqref{eq:tangent} therefore provides the consistent second-order decomposition.

\paragraph{Step 5: Non-negativity of the curvature term.}
Let $\mathcal C\in\mathbb R^{(4+n_a)\times8}$ be the matrix formed by stacking the rows of $\mathbf J^\ast$, $\nabla g(\mathbf z^\ast)^T$ (included if $\nu>0$), and $\mathbf e_j^T$ for $j\in\mathcal A_B$, where $n_a=|\mathcal A_B|+\mathbf 1_{\{\nu>0\}}$ denotes the number of active inequality constraints at $\mathbf z^\ast$. By Assumption~(A6), $\mathcal C$ has full row rank with minimum singular value $s_{\mathcal C}>0$, and $\ker\mathcal C$ forms the critical cone at $\mathbf z^\ast$. Suppose that the second-order sufficient condition holds---namely, $\mathbf u^T\mathbf H\mathbf u>0$ for all $\mathbf u\in\ker\mathcal C\setminus\{\mathbf 0\}$---and that $T_q=T_B=0$. Then $\nu g(\hat{\mathbf z})=0$, so $\nu>0$ implies $g(\hat{\mathbf z})=g(\mathbf z^\ast)=0$, while strict complementarity $\eta_j>0$ implies $b_j(\hat{\mathbf z})=|\delta_j|=0$ for all $j\in\mathcal A_B$. Combined with $\mathbf J^\ast\mathbf e_N=\mathbf 0$, this yields
\begin{equation*}
\begin{aligned}
\mathcal C\mathbf e_N&=\big(\mathbf 0;\,-\nabla g(\mathbf z^\ast)^T\mathbf s-R_g;\,(-s_j)_{j\in\mathcal A_B}\big),\\
\|\mathcal C\mathbf e_N\|&\le(G_q+1)\omega+\tfrac12L_gc_e^2\varepsilon^2=O(\varepsilon^2+\chi\varepsilon).
\end{aligned}
\end{equation*}
Projecting $\mathbf e_N$ onto $\ker\mathcal C$ via $\mathbf e_N^\circ=\mathbf e_N-\mathcal C^+\mathcal C\mathbf e_N\in\ker\mathcal C$ gives $\|\mathbf e_N-\mathbf e_N^\circ\|\le\|\mathcal C\mathbf e_N\|/s_{\mathcal C}$, whence
\begin{equation*}
\begin{aligned}
\mathbf e_N^T\mathbf H\mathbf e_N&=\mathbf e_N^{\circ T}\mathbf H\mathbf e_N^\circ+(\mathbf e_N-\mathbf e_N^\circ)^T\mathbf H(\mathbf e_N+\mathbf e_N^\circ)\\
&\ge-2\bar H\varepsilon\,\frac{\|\mathcal C\mathbf e_N\|}{s_{\mathcal C}}=-O(\varepsilon^3+\chi\varepsilon^2).
\end{aligned}
\end{equation*}
Thus, the curvature contribution $\tfrac12\mathbf e_N^T\mathbf H\mathbf e_N$ is non-negative up to $O(\varepsilon^3+\chi\varepsilon^2)$.

\paragraph{Remark on full-step acceptance.} The condition that the full step ($\alpha=1$) is accepted at the first iteration is readily verified in the local regime. For any trial step length $\alpha<1$, writing $\mathbf r_0+\alpha\mathbf A\mathbf d_0=(1-\alpha)\mathbf r_0+\alpha(\mathbf r_0+\mathbf A\mathbf d_0)$ and applying the estimates from the proof of Lemma~\ref{lem:contraction}---namely, $\|\mathbf r_0+\mathbf A\mathbf d_0\|\le(\bar\mu/\tilde\sigma)^2\|\mathbf r_0\|$, $\|(\mathbf J(\mathbf z_0)-\mathbf A)\mathbf d_0\|\le(\varepsilon_J/\tilde\sigma)\|\mathbf r_0\|$, and $\|\mathbf R_{r,0}\|\le\zeta_0\|\mathbf r_0\|$ with $\zeta_0=L\|\mathbf r_0\|/(2\tilde\sigma^2)$---yields $\|\mathbf r(\mathbf z_0+\alpha\mathbf d_0)\|\ge(1-\alpha-\chi-\zeta_0)\|\mathbf r_0\|$, whereas $\|\mathbf r(\mathbf z_0+\mathbf d_0)\|\le(\chi+\zeta_0)\|\mathbf r_0\|$. Consequently, every partial step produces a strictly larger residual norm than the full step whenever $1-\alpha>2(\chi+\zeta_0)$, a condition satisfied for all $\alpha\le\tfrac12$ as soon as $\chi+\zeta_0<\tfrac14$. Because the full step is inspected first in Algorithm~\ref{alg:svd} and achieves the smallest residual norm among the four candidates, it is the first candidate to reduce the residual and is therefore accepted (or returned directly upon meeting~\eqref{eq:accuracy} whenever the termination criterion is monotone in $\|\mathbf r\|$ across the candidate set).
\hfill$\square$

\bibliography{ast_references}

\begin{thebibliography}{35}
\providecommand{\natexlab}[1]{#1}

\bibitem[{Ba, Kiros, and Hinton(2016)}]{ba2016layernorm}
Ba, J.~L.; Kiros, J.~R.; and Hinton, G.~E. 2016.
\newblock Layer Normalization.
\newblock arXiv:1607.06450.

\bibitem[{Banerjee, Cauligi, and Pavone(2025)}]{banerjee2025iss}
Banerjee, S.; Cauligi, A.; and Pavone, M. 2025.
\newblock Deep Learning Warm Starts for Trajectory Optimization on the {International Space Station}.
\newblock In \emph{2025 International Conference on Space Robotics (iSpaRo)}, 394--401.

\bibitem[{Banerjee et~al.(2020)Banerjee, Lew, Bonalli, Alfaadhel, Alomar, Shageer, and Pavone}]{banerjee2020warmstart}
Banerjee, S.; Lew, T.; Bonalli, R.; Alfaadhel, A.; Alomar, I.~A.; Shageer, H.~M.; and Pavone, M. 2020.
\newblock Learning-based Warm-Starting for Fast Sequential Convex Programming and Trajectory Optimization.
\newblock In \emph{2020 IEEE Aerospace Conference}, 1--8.

\bibitem[{Benedikter et~al.(2021)Benedikter, Zavoli, Colasurdo, Pizzurro, and Cavallini}]{benedikter2021ascent}
Benedikter, B.; Zavoli, A.; Colasurdo, G.; Pizzurro, S.; and Cavallini, E. 2021.
\newblock Convex Approach to Three-Dimensional Launch Vehicle Ascent Trajectory Optimization.
\newblock \emph{Journal of Guidance, Control, and Dynamics}, 44(6): 1116--1131.

\bibitem[{Briden et~al.(2025)Briden, Choi, Yun, Linares, and Cauligi}]{briden2025constraint}
Briden, J.; Choi, C.; Yun, K.; Linares, R.; and Cauligi, A. 2025.
\newblock Constraint-Informed Learning for Warm-Starting Trajectory Optimization.
\newblock \emph{Journal of Guidance, Control, and Dynamics}, 48(10): 2272--2287.

\bibitem[{Chandler and Smith(1967)}]{chandler1967igm}
Chandler, D.~C.; and Smith, I.~E. 1967.
\newblock Development of the iterative guidance mode with its application to various vehicles and missions.
\newblock \emph{Journal of Spacecraft and Rockets}, 4(7): 898--903.

\bibitem[{D'Ambrosio, Silvestrini, and Lavagna(2026)}]{dambrosio2026sequence}
D'Ambrosio, M.; Silvestrini, S.; and Lavagna, M. 2026.
\newblock Conditioned Sequence Models for Warm-Starting Sequential Convex Trajectory Optimization in Space Robots.
\newblock \emph{Aerospace}, 13(2): 137.

\bibitem[{Dueri et~al.(2017)Dueri, A{\c{c}}{\i}kme{\c{s}}e, Scharf, and Harris}]{dueri2017customized}
Dueri, D.; A{\c{c}}{\i}kme{\c{s}}e, B.; Scharf, D.~P.; and Harris, M.~W. 2017.
\newblock Customized Real-Time Interior-Point Methods for Onboard Powered-Descent Guidance.
\newblock \emph{Journal of Guidance, Control, and Dynamics}, 40(2): 197--212.

\bibitem[{He et~al.(2016)He, Zhang, Ren, and Sun}]{he2016resnet}
He, K.; Zhang, X.; Ren, S.; and Sun, J. 2016.
\newblock Deep Residual Learning for Image Recognition.
\newblock In \emph{2016 IEEE Conference on Computer Vision and Pattern Recognition (CVPR)}, 770--778.

\bibitem[{Hendrycks and Gimpel(2016)}]{hendrycks2016gelu}
Hendrycks, D.; and Gimpel, K. 2016.
\newblock Gaussian Error Linear Units ({GELUs}).
\newblock arXiv:1606.08415.

\bibitem[{Horn and Johnson(2012)}]{horn2012matrix}
Horn, R.~A.; and Johnson, C.~R. 2012.
\newblock \emph{Matrix Analysis}.
\newblock Cambridge: Cambridge University Press, 2 edition.

\bibitem[{Jaggers(1977)}]{jaggers1977peg}
Jaggers, R.~F. 1977.
\newblock An explicit solution to the exoatmospheric powered flight guidance and trajectory optimization problem for rocket propelled vehicles.
\newblock In \emph{AIAA Guidance and Control Conference}. Hollywood, FL, USA.
\newblock {AIAA} Paper 77-1051.

\bibitem[{Knoll and Keyes(2004)}]{knoll2004jfnk}
Knoll, D.~A.; and Keyes, D.~E. 2004.
\newblock Jacobian-free {Newton--Krylov} Methods: A Survey of Approaches and Applications.
\newblock \emph{Journal of Computational Physics}, 193(2): 357--397.

\bibitem[{Lembono et~al.(2020)Lembono, Paolillo, Pignat, and Calinon}]{lembono2020memory}
Lembono, T.~S.; Paolillo, A.; Pignat, E.; and Calinon, S. 2020.
\newblock Memory of Motion for Warm-Starting Trajectory Optimization.
\newblock \emph{IEEE Robotics and Automation Letters}, 5(2): 2594--2601.

\bibitem[{Lentaris et~al.(2018)Lentaris, Maragos, Stratakos, Papadopoulos, Papanikolaou, Soudris, Lourakis, Zabulis, Gonzalez-Arjona, and Furano}]{lentaris2018hpec}
Lentaris, G.; Maragos, K.; Stratakos, I.; Papadopoulos, L.; Papanikolaou, O.; Soudris, D.; Lourakis, M.; Zabulis, X.; Gonzalez-Arjona, D.; and Furano, G. 2018.
\newblock High-Performance Embedded Computing in Space: Evaluation of Platforms for Vision-Based Navigation.
\newblock \emph{Journal of Aerospace Information Systems}, 15(4): 178--192.

\bibitem[{Levenberg(1944)}]{levenberg1944}
Levenberg, K. 1944.
\newblock A method for the solution of certain non-linear problems in least squares.
\newblock \emph{Quarterly of Applied Mathematics}, 2(2): 164--168.

\bibitem[{Li et~al.(2025)Li, Ran, Guo, Park, and Zhang}]{li2025marswarmstart}
Li, K.; Ran, G.; Guo, Y.; Park, J.~H.; and Zhang, Y. 2025.
\newblock Joint Trajectory Replanning for {Mars} Ascent Vehicle Under Propulsion System Faults: A Suboptimal Learning-Based Warm-Start Approach.
\newblock \emph{IEEE Transactions on Neural Networks and Learning Systems}, 36(12): 20302--20314.

\bibitem[{Liu, Lu, and Pan(2017)}]{liu2017survey}
Liu, X.; Lu, P.; and Pan, B. 2017.
\newblock Survey of convex optimization for aerospace applications.
\newblock \emph{Astrodynamics}, 1(1): 23--40.

\bibitem[{Lu, Zhang, and Hu(2024)}]{initialcostates2024}
Lu, S.; Zhang, Y.; and Hu, Q. 2024.
\newblock Initial costates derived by near-optimal reference sequence and least-squares method.
\newblock \emph{Chinese Journal of Aeronautics}, 37(5): 377--391.

\bibitem[{Ma et~al.(2022)Ma, Pan, Hao, and Tang}]{ma2022mcpi}
Ma, Y.; Pan, B.; Hao, C.; and Tang, S. 2022.
\newblock Improved sequential convex programming using modified {Chebyshev--Picard} iteration for ascent trajectory optimization.
\newblock \emph{Aerospace Science and Technology}, 120: 107234.

\bibitem[{Ma et~al.(2025)Ma, Pan, Huang, Chen, and Xu}]{reducedspace2025}
Ma, Y.; Pan, B.; Huang, L.; Chen, Q.; and Xu, Z. 2025.
\newblock High-efficiency reduced space sequential convex programming using low-complexity matrix inversion.
\newblock \emph{Aerospace Science and Technology}, 159: 109958.

\bibitem[{Malyuta et~al.(2021)Malyuta, Yu, Elango, and A{\c{c}}{\i}kme{\c{s}}e}]{malyuta2021advances}
Malyuta, D.; Yu, Y.; Elango, P.; and A{\c{c}}{\i}kme{\c{s}}e, B. 2021.
\newblock Advances in trajectory optimization for space vehicle control.
\newblock \emph{Annual Reviews in Control}, 52: 282--315.

\bibitem[{Marquardt(1963)}]{marquardt1963}
Marquardt, D.~W. 1963.
\newblock An Algorithm for Least-Squares Estimation of Nonlinear Parameters.
\newblock \emph{Journal of the Society for Industrial and Applied Mathematics}, 11(2): 431--441.

\bibitem[{Rogers, Barnhart, and Crago(2016)}]{rogers2016maestro}
Rogers, C.~M.; Barnhart, D.; and Crago, S. 2016.
\newblock The {Maestro} Flight Experiment: A 49-Core Radiation Hardened Processor in Space.
\newblock In \emph{2016 IEEE Aerospace Conference}, 1--6.

\bibitem[{Rubinsztejn, Sood, and Laipert(2020)}]{rubinsztejn2020missedthrust}
Rubinsztejn, A.; Sood, R.; and Laipert, F.~E. 2020.
\newblock Neural network optimal control in astrodynamics: Application to the missed thrust problem.
\newblock \emph{Acta Astronautica}, 176: 192--203.

\bibitem[{Saad and Schultz(1986)}]{saad1986gmres}
Saad, Y.; and Schultz, M.~H. 1986.
\newblock {GMRES}: A Generalized Minimal Residual Algorithm for Solving Nonsymmetric Linear Systems.
\newblock \emph{SIAM Journal on Scientific and Statistical Computing}, 7(3): 856--869.

\bibitem[{S{\'a}nchez-S{\'a}nchez and Izzo(2018)}]{sanchez2018landing}
S{\'a}nchez-S{\'a}nchez, C.; and Izzo, D. 2018.
\newblock Real-Time Optimal Control via Deep Neural Networks: Study on Landing Problems.
\newblock \emph{Journal of Guidance, Control, and Dynamics}, 41(5): 1122--1135.

\bibitem[{Shi et~al.(2022)Shi, Wang, Su, Ma, and Chen}]{shi2022indirect}
Shi, J.; Wang, J.; Su, L.; Ma, Z.; and Chen, H. 2022.
\newblock A Neural Network Warm-Started Indirect Trajectory Optimization Method.
\newblock \emph{Aerospace}, 9(8): 435.

\bibitem[{Takubo et~al.(2026)Takubo, Adang, Schwager, and D'Amico}]{takubo2026transformer}
Takubo, Y.; Adang, M.; Schwager, M.; and D'Amico, S. 2026.
\newblock Transformer-Based Warm-Starting for Feasible and Optimal Terminal Approach to Tumbling Objects with Space Manipulators.
\newblock arXiv:2606.17317.

\bibitem[{Tambara et~al.(2019)Tambara, Hernandez, Sturesson, Hjorth, Andersson, and Weigand}]{tambara2019gr740}
Tambara, L.~A.; Hernandez, F.; Sturesson, F.; Hjorth, M.; Andersson, J.; and Weigand, R. 2019.
\newblock Single Event Effect Characterization of the {GR740} Rad-Hard Quad-Core {LEON4FT} System-on-Chip.
\newblock In \emph{2019 19th European Conference on Radiation and Its Effects on Components and Systems (RADECS)}, 1--6.

\bibitem[{Virtanen et~al.(2020)Virtanen, Gommers, Oliphant et~al.}]{virtanen2020scipy}
Virtanen, P.; Gommers, R.; Oliphant, T.~E.; et~al. 2020.
\newblock {SciPy} 1.0: Fundamental algorithms for scientific computing in {Python}.
\newblock \emph{Nature Methods}, 17(3): 261--272.

\bibitem[{Wang et~al.(2018)Wang, Lo, Hwang, and Hwang}]{wang2018fqlnk}
Wang, H.-H.; Lo, Y.-S.; Hwang, F.-T.; and Hwang, F.-N. 2018.
\newblock A Full-Space Quasi-{Lagrange--Newton--Krylov} Algorithm for Trajectory Optimization Problems.
\newblock \emph{Electronic Transactions on Numerical Analysis}, 49: 103--125.

\bibitem[{Wang et~al.(2022)Wang, Dai, Cheng, Liu, Cui, Zhang, and Feng}]{wang2022ascent}
Wang, X.; Dai, P.; Cheng, X.; Liu, Y.; Cui, J.; Zhang, L.; and Feng, D. 2022.
\newblock An online generation method of ascent trajectory based on feedforward neural networks.
\newblock \emph{Aerospace Science and Technology}, 128: 107739.

\bibitem[{Yun et~al.(2025)Yun, Bayard, Kubiak, Owens, Johnson, Johnson, Scharf, and Lu}]{yun2025multicore}
Yun, K.; Bayard, D.; Kubiak, G.; Owens, A.; Johnson, A.; Johnson, R.; Scharf, D.; and Lu, T. 2025.
\newblock Enhancing Fault-Tolerant Space Computing: Guidance Navigation and Control ({GNC}) and Landing Vision System ({LVS}) Implementations on Next-Gen Multi-Core Processors.
\newblock In \emph{2025 IEEE Space Computing Conference (SCC)}, 29--34.

\bibitem[{Zhou et~al.(2026)Zhou, Zhang, Wang, Xu, Sun, Qin, and Gu}]{fixedwing2026}
Zhou, Y.; Zhang, C.; Wang, Y.; Xu, G.; Sun, Q.; Qin, Q.; and Gu, L. 2026.
\newblock Trajectory optimization learning for fixed-wing aircraft.
\newblock \emph{Chinese Journal of Aeronautics}, 104388.

\end{thebibliography}

\end{document}